\documentstyle[aps,pre,fleqn,epsf,floats]{revtex}
\begin{document}
\def\theequation{\arabic{section}.\arabic{equation}}
\global\firstfigfalse


\title{UNIFIED MODEL FOR THE STUDY OF DIFFUSION 
LOCALIZATION AND DISSIPATION}

\author{Doron Cohen}

\date
{
August 1996 \footnote{Published in Phys. Rev. {E \bf 55}, 1422-1441 (1997).}    
}

\address
{
Department of Physics of Complex Systems, \\ 
The Weizmann Institute of Science, Rehovot 76100, Israel.
} 

\maketitle


\begin{abstract}
A new model that generalizes the study of quantum Brownian 
motion (BM) is constructed. We consider disordered environment that 
may be either static (quenched), noisy or dynamical. 
The Zwanzig-Caldeira-Leggett BM-model constitutes formally 
a special case where the disorder auto-correlation 
length is taken to be infinite. Alternatively, localization problem 
is obtained if the noise auto-correlation time is taken to be infinite. 
Also the general case of weak nonlinear coupling to thermal, 
possibly chaotic bath is handled by the same formalism.  
A general, Feynman-Vernon type path-integral expression for the
propagator is introduced. Wigner transformed version of this expression 
is utilized in order to facilitate comparison with the classical limit.
It is demonstrated that non-stochastic genuine quantal manifestations 
are associated with the new model. It is clarified that such effects 
are absent in the standard BM model, either the disorder or the chaotic 
nature of the bath are essential. Quantal correction to the classical 
diffusive behavior is found even in the limit of high temperatures.  
The suppression of interference due to dephasing is discussed, leading 
to the observation that due to the disorder the decay of coherence is 
exponential in time, and no longer depends on geometrical considerations.  
Fascinating non-Markovian effects due to time-correlated (colored) noise 
are explored. For this, a new strategy is developed in order to 
handle the integration over paths. This strategy is extended in 
order to demonstrate how localization comes out from the path integral 
expression. 
\end{abstract}


\section{Introduction}
\setcounter{equation}{0}

The dynamics of a particle that interacts with its 
environment constitutes a basic problem in physics.
Classically, upon elimination of the environmental 
degrees of freedom, the reduced dynamics is most
simply described in terms of Langevin equation. 
Solution of this equation, by utilizing Fokker Planck
equation, is well known \cite{krm}. In the absence 
of external potential, it yields spreading and 
diffusion. The latter effect is due to the interplay
of noise and dissipation. However, diffusion may arise 
also from the interaction with disordered environment.
This kind of non-dissipative ``random-walk'' diffusion
is encountered, for example, in Solid State Physics, while
analyzing electrical conductivity. 
It is well known \cite{mott} that this latter type of 
diffusion may be suppressed quantum mechanically due 
to localization effect. Still, diffusive-like behavior 
is recovered if noise and dissipation are taken into 
account.  
  
The unified modeling of the environment in terms of 
noise, dissipation and disorder is the 
first stage of the present study.
One may take the notion of particle literally, 
and identify the environment as either external 
or internal bath that consists of infinitely many
degrees of freedom.  The bath may be a large 
collection of other particles or field modes
(photons, phonons). Else, the internal degrees of freedom  
of the particle itself are considered to be the 
bath. The latter point of view has been suggested
by Gross \cite{gross} in order to analyze inelastic  
scattering of heavy ions.  

A totally different point of view, promoted by 
Caldeira and Leggett (CL) \cite{CL}, 
consider the notion of particle as 
a token for some macroscopic degree of freedom.
A linear interaction with a speculated 
bath that consists of infinitely many uncoupled harmonic 
oscillators, is assumed (Zwanzig \cite{gross}). 
The known classical 
limit, namely Langevin-type equation, serves as 
a guide for the construction of the proper 
Hamiltonian. 
(Phenomenological rather than microscopic considerations
are used, hence the usage of the term `speculated bath').
The power inherent in this  
approach is the capability to introduce an 
explicit path integral expression for the
propagator, using the Feynman Vernon (FV) \cite{FV}
formalism. This FV-CL propagator constitutes 
a quantized description of Brownian motion (BM).
The term ``BM-model'' will be associated from now on 
with this propagator.

The first question which should be asked concerning
the applicability of CL approach is obviously whether
either the coupling, the bath, or both are ``too
simple'' in order to account for the variety of 
physical phenomena that are associated with generalized BM.
The term ``generalized BM'' is used in order to describe 
any dynamical behavior that corresponds in the classical limit
to Langevin-like equation. 
In the simplest case Langevin equation is 
$m\ddot{x}+\eta\dot{x}={\cal F}$ where $m$ and $\eta$
are the mass and the friction coefficient, respectively. 
One should specify the stochastic force ${\cal F}$. 
This force may arise from interaction with some 
fluctuating homogeneous field 
(${\cal F}={\cal F}(t)$), which is the  
usual formulation. However, {\em more generally}, this force may 
arise from the interaction with {\em disordered} 
potential (${\cal F}=-\nabla{\cal U}(x,t)$).
In the latter case the spatial auto-correlations 
of the force are significant.
To avoid misunderstanding it should be emphasized
that there are other aspects in which BM can be 
generalized (for review see \cite{hanggi}).  

In the present paper we construct a unified model for the 
study of Diffusion, Localization and Dissipation (DLD). 
This model describes generalized BM in the sense 
specified above.
The disordered environment may be either 
static (quenched), noisy or dynamical. 
The model is treated within the framework of the FV 
formalism. The resultant path integral expression for the 
propagator contains a functional $S_F$ with kernel 
$\alpha(\tau{-}\tau')$ that corresponds to friction, 
and a functional $S_N$ with kernel $\phi(\tau{-}\tau')$ that 
corresponds to the noise.  Both functionals depend also on a 
suitably defined auto-correlation function $w(r{-}r')$ that 
characterizes the disorder.   
BM model constitutes (formally) a special case where the 
disorder auto-correlation length is taken to be infinite. 
Optional derivations of the resultant path integral 
expression are presented for the particular cases of either classical 
or quantal systems with non-dynamical disordered potential.
Obviously, in the latter case the friction functional $S_F$ is not 
generated by the derivation. Localization problem is obtained 
if the noise auto-correlation time is taken to be infinite 
($\phi(\tau{-}\tau')=const$).
  
An explicit computation of both the classical and the 
quantal propagators ${\cal K}(R,P|R_0,P_0)$ will be carried out. 
This propagator generates the time evolution of $\rho(R,P)$,
which is either the Wigner function or the corresponding 
classical phase-space distribution. Spreading and diffusion profiles 
are found for either noisy or ohmic environment. Quantal corrections 
to the classical result are discussed.  A new strategy is developed 
in order to handle the integration over paths. This strategy is 
utilized in order to study the anomalous ``diffusion'' profiles 
due to colored noise. Later it is extended in order to demonstrate 
how localization comes out from the path integral expression.

Again, one may ask, whether the DLD model is 
the ``ultimate'' model for the description of BM
in the most generalized way (as far as generic 
effects are concerned). 
In case of 2-D generalized BM one should consider 
also the effect of ``geometric magnetism'' \cite{berry},
which is not covered by the 1-D DLD model.
Here we limit the discussion to 1-D BM.  
In order to answer this 
question one should consider a general 
nonlinear coupling to a thermal, possibly 
chaotic bath. In the limit of weak coupling  
one may demonstrate (see Sec.VI(B)) that indeed
the bath can be replaced by an equivalent 
``effective bath'' that consists of harmonic 
oscillators, yielding the DLD model.
A further reduction to BM model is achieved
if the coupling is linear. The derivation 
also demonstrates why the so called ``ohmic''
bath is generic.  However, we cannot prove 
that the path-integral expression that corresponds 
to the DLD model is the most generalized description
of BM 
(in the sense of this paper). 
Gefen and Thouless \cite{efrat}, Wilkinson \cite{wilk} 
and Shimshoni and Gefen \cite{efrat} have emphasized 
the significance of Landau-Zener transitions as a
mechanism for dissipation. The weak coupling
approximation misses this effect.  Still, 
there is a possibility that some future, more 
sophisticated derivation, will demonstrate 
that an equivalent ``oscillators bath'' can 
be defined also in the case of strong coupling.
The existence of such derivation is most 
significant, since it implies that no ``new effects''
(such as ``geometric magnetism'' in case of 2-D 
generalized BM) can be found in the context of 
1-D generalized BM.
Referring again to the Landau-Zener mechanism,
Wilkinson has demonstrated that anomalous
friction, which is not proportional to velocity,
may arise \cite{wilk}. 
The BM model cannot generate such anomalous effect. However, 
we shall demonstrate that the {\em non-ohmic} DLD model
can be used in order to generate this effect.

We turn to review previous works that are related
to the present study.  Traditional approach
to the study of dissipation is based on the  
Master Equation formalism \cite{louisell}, which 
is the quantal analog of the classical Fokker Planck 
treatment \cite{krm}. Systematic derivations
are usually based, in some stage, on the Markovian  
approximation.  An alternative route is to apply 
the FV formalism \cite{FV}. This has been done by M\"{o}hring
and Smilansky \cite{uzy}, following 
Gross \cite{gross}, in order to study deep inelastic collisions 
of heavy ions. Later, the FV formalism has been 
applied by CL \cite{CL} 
and followers \cite{hanggi} in particular to the study of 
macroscopic quantum tunneling. 
Hakim and Ambegaokar \cite{HA} has applied FV model 
in order to compute the spreading and the diffusion 
of quantum Brownian particle. Cohen and Fishman \cite{CF} 
have computed the full Wigner propagator 
in case of general quadratic Hamiltonian, possibly 
time dependent. The latter study has demonstrated  
more clearly
the significance of {\em noise time-autocorrelations}.
Non-trivial auto-correlations may lead to non-Markovian 
effects due to the non-local (in time) nature of the 
noise functional $S_N$. The simplest example is the 
suppression of diffusion at low temperatures \cite{HA}
due to negative power-law correlations of the noise \cite{CF}. 
A less trivial manifestation of non-Markovian effect has been   
found by Cohen \cite{noise} while analyzing diffusion due to 
the destruction of localization by colored noise.

Non Markovian features are usually speculated 
to be less relevant if disorder is taken into 
account. One may expect that for generalized BM, 
disorder will lead, at all temperatures, 
to normal diffusion. We shall find later in this 
paper that non-Markovian effects are quite 
effective also in case of the DLD model, 
and lead to anomalous spreading profiles. 
However, it will be clear that these effects, 
though counter intuitive at first sight, are 
of classical nature.  Non Markovian effects can also 
arise from the non-locality of the friction 
functional $S_F$, this is the case for non-ohmic
bath. The retarded response of the bath has then 
long memory for the particle's dynamics, see for
example the review papers by Grabert et al. and by 
H\"{a}nggi et al. \cite{hanggi}.  
In particular, for unbounded motion, in the absence 
of disorder (BM model), it will be demonstrated that one 
encounters infinities in computations of the friction and 
of the effective mass. These infinities are 
avoided if disorder is taken into account (Sec.IV(B)).

The suppression of quantal interference due to 
dephasing process is an important issue for both 
semiclassical \cite{uzy} and mesoscopic \cite{stern} physics.
The dephasing of interference in metals due to 
electromagnetic fluctuations has been discussed by 
Al'tshuler, Aronov and Khmelnitskii \cite{AAK}.
General considerations has been presented 
by Stern, Aharonov and Imry \cite{stern}. 
The strong dimensionality dependence of 
the dephasing process has been emphasized.
In this paper, the DLD model is used in order to study 
the suppression of quantal interference due to the local
interaction with either noisy or dynamical {\em disordered} 
environment. The dephasing  is determined by the noise functional 
$S_N[x''(\tau),x'(\tau)]$.  Due to the local nature of the 
dephasing process, the decay of coherence is exponential in time, 
and no longer depends on geometrical considerations.

In case of quenched disordered environment the DLD
model reduces to a localization problem that its 
solution is well known \cite{lifshitz}.  In particular, 
for $\delta$ correlated potential the result for 
the localization length has been pointed out by 
Thouless \cite{thouless}. As far as we know, the real-time 
Feynman path integral formalism has not been utilized 
so far in order to re-derive this result, though 
functional integration is frequently employed 
in closely related computations.  The case of 
noisy disordered environment, where the potential is
$\delta$ correlated both in time and in space has 
been considered by Jayannavar and Kumar \cite{kumar}.  In the latter 
reference the spatial spreading has been computed, and 
a classical-like result has been obtained. Quantal 
corrections to the dispersion profile, has not
been discussed in the latter reference.
No solution exists for a model that ``interpolates'' the 
crossover from quenched noise, via colored noise, to white noise 
disordered potential.
Marianer and Deutsch \cite{marianer} have considered the 
problem of white noisy disordered potential with added dissipation. 
Using BM model, they have demonstrated that a classical-like results 
is obtained for the spatial spreading. Again, neither the spreading 
profile nor quantal corrections have been considered.

The outline of this paper is as follows: \ 
In Sec.II the unified model for the study of Diffusion, 
Localization and Dissipation (DLD) is constructed. Derivation
of the reduced classical dynamics is presented in Sec.III-IV. 
It is shown that a well defined Langevin equation is obtained 
for both subohmic and superohmic bath, as well as for ohmic bath. 
This is contrasted with BM model, where only  the ohmic case is 
well defined. 
In Sec.V-VI, four derivations of the FV path integral 
expression for the propagator are presented: (a) A classical 
derivation that is based on Langevin equation; (b) A quantum mechanical 
derivation for non-dynamical noisy or quenched disordered potential; 
(c) A quantum mechanical derivation for the full DLD model;  
(d) A quantal derivation for the general case of weak nonlinear
coupling to a thermal, possibly chaotic bath. 
App.A clarifies the relation of BM-model and DLD-model
to the case of interaction with external bath that consists of 
extended field modes. In Sec.VI(C) the explicit expression 
for the influence functional allows concrete predictions  
concerning the loss of interference due to dephasing.
App.B introduce a gedanken experiment that clarifies the manifestation 
of interference and dephasing in the presence of dynamical disordered 
environment.   
In Sec.VII-VIII three strategies for the computation of the 
quantal propagator are discussed. Spreading and diffusion profiles are
found for either noisy or ohmic environment. The DLD model is compared 
with the BM model. In the latter case the result is classical-like, 
while in case of the DLD model, a singular ``quantal correction'' should 
be included. The significance of this ``correction'' is further 
clarified in App.B.  
The strategy of Sec.VIII, which enables computation of diffusion 
profiles in the presence of {\em colored} noise, is 
extended in Sec.IX. We use this strategy in order to demonstrate 
how, in the case of {\em quenched} noise, localization comes out from the 
general FV path integral expression. Summary and Conclusions are 
presented in Sec.X.


\section{Construction of the DLD Model}
\setcounter{equation}{0}

\subsection{Langevin Equation}  

As a starting point for later generalization 
we consider the classical Langevin equation.  
This equation describes the time evolution of 
Brownian particle under the the influence of 
so called ``ohmic'' friction and stochastic force. 
\begin{eqnarray}   \label{e2_1}
m\ddot{x}+\eta\dot{x}+V'(x)={\cal F}  
\ \ \ \ \mbox{[Langevin Equation]} \ \ \ \ .
\end{eqnarray}
In the {\em standard} Langevin equation the stochastic 
force represents stationary ``noise'' which is zero 
upon ensemble average, and whose autocorrelation  
function is 
\begin{eqnarray}   \label{e2_2}
\langle{\cal F}(t) {\cal F}(t')\rangle = \phi(t{-}t')
\ \ \ \ \mbox{[for standard Langevin]}. 
\end{eqnarray}
Usually white noise, which is $\delta$ correlated 
in time, is considered. The standard Langevin equation
can be {\em generalized} by assuming that the stochastic 
force is due to some noisy potential, namely 
$ {\cal F}(x,t) = -{\cal U}'(x,t) $
where $\prime$ denotes spatial derivative. 
One may assume that ${\cal U}$ is zero upon 
ensemble averaging, and satisfies 
\begin{eqnarray} \label{e2_3} 
\langle{\cal U}(x,t) {\cal U}(x',t')\rangle = \phi(x{-}x',t{-}t')       
\ \ \ \ \mbox{[generalized Langevin]}
\end{eqnarray}
The autocorrelation function of the stochastic 
force ${\cal F}$ at some specified point $x$ is 
$ \phi(t{-}t') \equiv - \phi''(0,t{-}t')  $
In practice $\phi(x{-}x',t{-}t')$ will be assumed 
to have the factorized form 
\begin{eqnarray}  \label{e2_4}
\phi(x{-}x',t{-}t') = \phi(t{-}t') \cdot w(x{-}x') 
\end{eqnarray}
where, without loss of generality, $w(r)$ is normalized so that
$w''(0)=-1$, the normalization constant being absorbed 
into $\phi(\tau)$. 

For distribution of particles, the {\em standard} Langevin equation 
predicts rigid motion with no diffusion. This feature is eliminated 
if an average over realizations of ${\cal F}$ is performed.  
The {\em standard} Langevin equation can be viewed as a special 
case of the {\em generalized} version (\ref{e2_3}). For this one 
should take $ {\cal U}(x,t) = -x \cdot {\cal F}(t) $. 
Alternatively,
\begin{eqnarray} \label{e2_5}
w(x-x') \ = \ -\frac{1}{2}(x-x')^2 
\ \ \ \ \ \mbox{[for standard Langevin]}.
\end{eqnarray}  
Another choice for spatial autocorrelations that
corresponds to disordered environment is
\begin{eqnarray}  \label{e2_6}
w(x-x')=\ \ell^2\exp
\left(-\frac{1}{2}\left(\frac{x-x'}{\ell}\right)^2\right)
\ \ \ \ \mbox{[for disordered environment]} .
\end{eqnarray}
The parameter $\ell$ constitutes a measure for the microscopic
scale of the disorder. For distribution of particles, the 
resultant motion will be diffusive-like rather than rigid, 
even without averaging over realizations. It is crucial 
to do this important observation if one wishes to introduce  
a {\em quantized} version of Langevin equation. 

 
\subsection{Hamiltonian formulation} 

Disregarding the friction term, the Langevin equation may be derived
from the Hamiltonian
\begin{eqnarray}      \label{e3_1}
{\cal H} = \frac{p^2}{2m} + V(x) + {\cal U}(x,t)
\end{eqnarray}
where average over realizations of ${\cal U}$ is implicit. If 
${\cal U}$ is time independent, we shall use the notion {\em quenched}
disordered environment. If ${\cal U}$ is uncorrelated 
also in time, we shall use the notion {\em noisy} environment. 

If $\eta$ is non-zero, energy may dissipate, and we should  
consider ``dynamical'' environment rather than ``noisy'' or ``quenched'' one.
In order to model the environment we generalize the FV-CL approach. 
Namely,  to facilitate later mathematical treatment, the interaction 
potential is assumed to be linear in the bath coordinates:  
\begin{eqnarray}     \label{e3_2}
{\cal U}(x,t) \ \ \rightarrow \ \  {\cal H}_{int} =
 - \sum_{\alpha} c_{\alpha} Q_{\alpha} u_{\alpha} (x) 
\end{eqnarray}
where $ Q_{\alpha} $ denotes the dynamical coordinate of the $\alpha$
scatterer or bath mode. $u_{\alpha} (x)$ is the normalized interaction  
potential, and $c_{\alpha}$ are coupling constants. 
In case of dynamical environment, the $Q_{\alpha}$ are assumed 
to be oscillators' coordinates. The bath Hamiltonian is:
\begin{eqnarray}   \label{e3_3}
{\cal H}_{bath}=\sum_{\alpha}(\frac{P_{\alpha}^2}{2m_{\alpha}}
+\frac{1}{2} m \omega_{\alpha}^2 Q_{\alpha}^2) \ \ \ \ .
\end{eqnarray}
The total Hamiltonian of the system plus the heat bath is
\begin{eqnarray}  \label{e3_4}
{\cal H} = \frac{p^2}{2m} + V(x) + {\cal H}_{int} 
+ {\cal H}_{bath} \ \ \ \ . 
\end{eqnarray} 
In order to further specify the DLD model, we should characterize 
the spectral distribution of the bath-oscillators, as well as
their interaction with the particle \cite{eli}. 

In what follows we consider the case of localized 
scatterers. The scatterers are assumed to be uniformly 
distributed in space with $u_{\alpha}(x)=u(x{-}x_{\alpha})$. 
The joint distribution of the bath oscillators with respect 
to $\omega_{\alpha}$ and $x_{\alpha}$ is assumed to be 
factorized. Consequently we can write    
\begin{eqnarray}    \label{e3_5}
\frac{\pi}{2} \sum_{\alpha}
\frac{c^2_{\alpha}}{m_{\alpha}\omega_{\alpha}} 
\delta(\omega-\omega_{\alpha}) \ \delta(x-x_{\alpha})
\ = \ J(\omega)  \ \ \ \ .
\end{eqnarray}
The bath is characterized by the spectral function
$J(\omega)$. If we consider a partition of the bath 
oscillators into subsets of oscillators whose  
positions $x_{\alpha}$ are the same, then locally, within 
each subset, the $\omega_{\alpha}$ will have 
the same distribution.  
The interaction is characterized by well defined spatial 
autocorrelation function $w(x{-}x')$, namely 
\begin{eqnarray}  \label{e3_6}
w(r) \ = \ \int_{-\infty}^{\infty} u(r-x') u(x') dx'
\ = \ \frac{1}{\mbox{density}} 
\sum_{\alpha} u_{\alpha}(R+r) u_{\alpha}(R) 
\ \ \ \ , 
\end{eqnarray}
where ``density'' refers to the uniform 
spatial distribution $\sum\delta(x{-}x_{\alpha})$.
The scattering potential $u(x)$ will be normalized 
so that $w''(0)=-1$, the normalization constant being 
absorbed in the coefficients $c_{\alpha}$. 
Disregarding for a moment the dynamical nature of the bath, 
thus considering again the case of either ``noisy'' or 
``quenched'' environment, one obtains  
$\langle{\cal U}(x,t) {\cal U}(x',t')\rangle=
\sum_{\alpha} c_{\alpha}^2 
\langle Q_{\alpha}(t) Q_{\alpha}(t')\rangle
u_{\alpha}(x) u_{\alpha}(x') $
which upon recalling previous definitions, 
leads to (\ref{e2_4}) with
$\phi(t{-}t') = \mbox{density}\times\sum_{\alpha} c_{\alpha}^2 
\langle Q_{\alpha}(t) Q_{\alpha}(t')\rangle $ \ .
Thus, if the dynamical nature of the bath is 
ignored, the problem reduces to solving 
Langevin equation with the appropriate $w(x{-}x')$ 
and $\phi(t{-}t')$. 

The spectral function $J(\omega)$ and the 
spatial autocorrelation function $w(r)$ 
constitute a complete specification of the 
model Hamiltonian (\ref{e3_4}). This Hamiltonian   
will be used in order to study diffusion  
localization and dissipation (DLD) within the 
framework of a unified formalism.   

The BM model, where $u_{\alpha}(x)=x$ up to normalization, 
constitutes {\em formally} a special case of the DLD model. 
It will be possible to 
apply the ``unified'' formalism that will be developed 
for the DLD model, also for the analysis of the 
BM model, merely by substitution of the 
appropriate $w(r)$. Namely, one may use $w(r)$ 
of (\ref{e2_5}), or alternatively  one may take the 
limit $\ell\rightarrow\infty$ in (\ref{e2_6}). However, 
there are some minor, but significant differences  
that will be noted in due time. This is because
our derivations rely on the {\em local} nature 
of $w(r)$. Therefore we consider the notion 
``BM model'' distinctively from the notion ``DLD model''. 
Also the results will be both qualitatively and 
quantitatively different, in-spite of being  
derived, formally, from look-alike formulas.

\begin{figure} 
\begin{center}
\leavevmode 
\epsfysize=8in
\epsffile{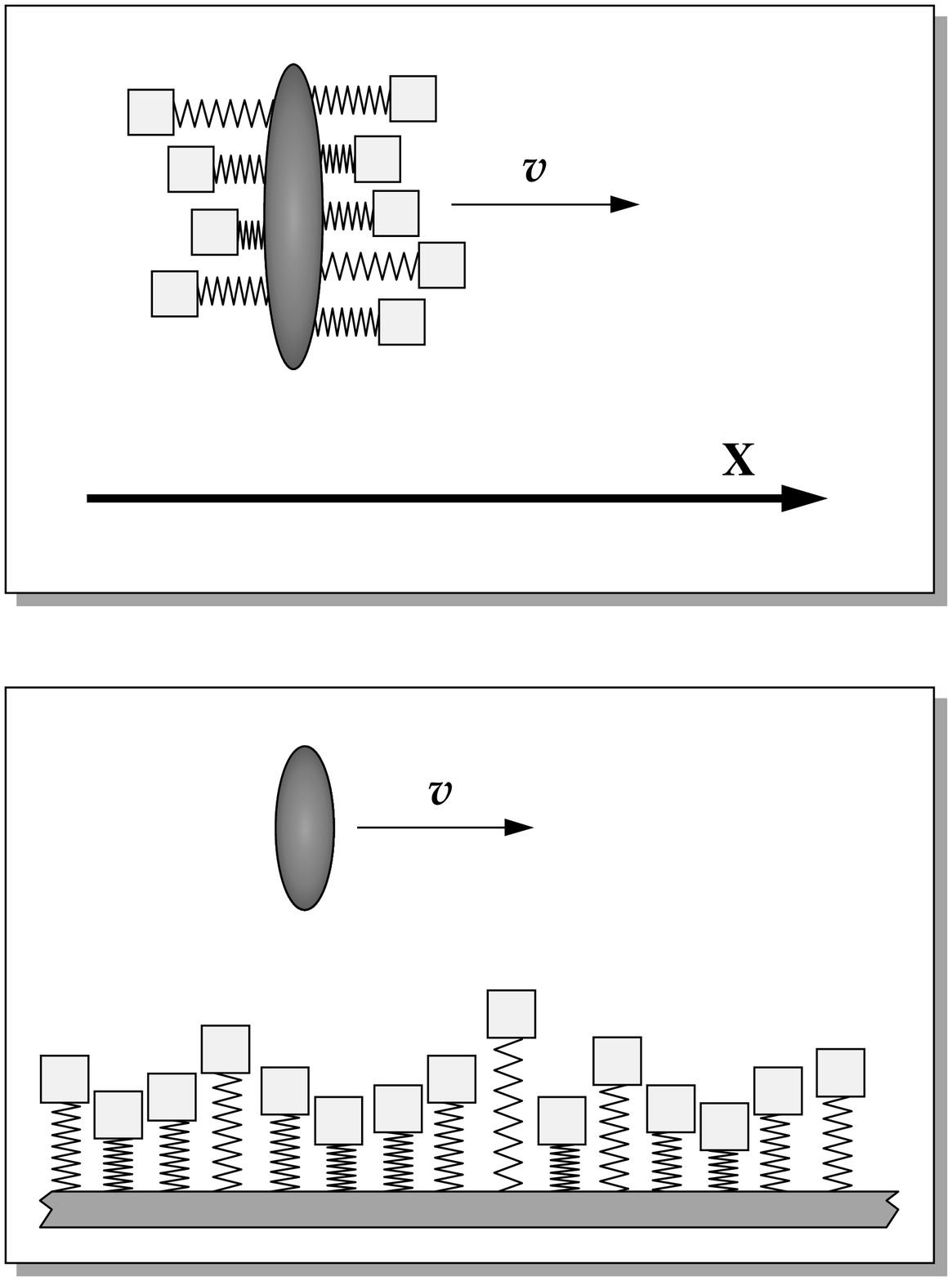}
\end{center}
\caption{\protect\footnotesize Illustration of the BM 
 model (top drawing), versus the DLD model (lower drawing). 
 Note that in the first case a counter term has been 
 incorporated as in (\protect\ref{e_counter})}
\label{models}
\end{figure}

Fig.\ref{models} illustrates the BM Hamiltonian and the 
DLD Hamiltonian.
Looking at the figure it is apparent that the natural 
application of the BM model is for the description 
of the dynamics of a composite particle, with many internal 
degrees of freedom. This interpretation is due to Gross 
Ref.\cite{gross}, who suggested to apply such models for 
the analysis of heavy-ion collisions. The bath according 
to this interpretation is ``internal'', carried by the 
particle, rather than ``external''.  

The interaction Hamiltonian (\ref{e3_2}) may also describe 
the interaction with some ``external'' bath that 
consists of extended field modes. For example,  
the electron-photon interaction and the electron-phonon 
interaction may be cast into the form of (\ref{e3_2}), with 
$u_{\alpha}(x)=\cos(\omega_{\alpha}/c \cdot x), \ 
\sin(\omega_{\alpha}/c \cdot x)$ , where $c$ is either 
the speed of light or the speed of sound. In the long wavelength 
limit this interaction resembles the BM-model rather 
than the DLD-model. With minor modifications, also these 
models may be treated within the framework of our 
``unified'' treatment (Appendix A).  

Finally, we re-emphasize that in general, the DLD model 
as well as the BM model may be used on a phenomenological 
basis for the description of dissipation in mesoscopic 
quantum devices.  This point of view will be discussed 
further later on (Sec.VI(B) in particular).

 
\section{Derivation of Langevin Equation} 
\setcounter{equation}{0}

\subsection{The Reduced Equation of motion} 

The quantal state of the particle may be represented by Wigner
function $\rho (R,P)$. \ The time evolution of Wigner function
corresponds to that of classical distribution in phase-space.  
In this section we consider a `classical treatment of the dynamics'. 
The latter term implies that the system is considered to be classical, 
($\hbar_{system}\rightarrow 0$), while the bath gets full quantum 
mechanical treatment. The limit $\hbar_{bath} \rightarrow 0$ is not 
taken. The equations of motion of classical points that form a 
distribution in phase-space are 
\mbox{$ \dot{x} = p $} and 
\begin{eqnarray}  \label{e4_1}
\dot{p} = - V^{\prime}(x) + {\cal F}_{bath}  
\ \ , \ \ \ \ \
{\cal F}_{bath} = \sum_{\alpha} c_{\alpha} Q_{\alpha}(t) 
u^{\prime}_{\alpha}(x) \ \ \ \ . 
\end{eqnarray}
The variables $Q_{\alpha}(t)$ satisfy the equation
\begin{eqnarray}
m_{\alpha} \ddot{Q}_{\alpha}(t) + m_{\alpha} \omega^2_{\alpha}
Q_{\alpha}(t) = c_{\alpha} u_{\alpha}(x)
\nonumber 
\end{eqnarray}                                        
which can be solved explicitly, namely,
\begin{eqnarray}
Q_{\alpha}(t) = Q_{\alpha}(0) \cos(\omega_{\alpha}t) +
\frac{P_{\alpha}(0)}{m_{\alpha}\omega_{\alpha}} \sin (\omega_{\alpha} t)
+ \int_0^t dt^{\prime} \frac{c_{\alpha}}{m_{\alpha}\omega_{\alpha}}
\sin(\omega_{\alpha}(t-t^{\prime})) u_{\alpha}(x(t^{\prime})) \ .
\nonumber
\end{eqnarray}
Substitution of the latter expression into (\ref{e4_1}) yields
\begin{eqnarray}  \label{e4_2}
{\cal F}_{bath} =  {\cal F}_{retarded} + {\cal F}_{noise}
\end{eqnarray}
with
\begin{eqnarray}   \label{e4_3}
{\cal F}_{retarded} = \int_0^t dt^{\prime} \ 
2\alpha(t{-}t^{\prime}) \ w^{\prime}(x(t){-}x(t^{\prime}))
\ \ \ \ .
\end{eqnarray} 
The response kernel $\alpha(\tau)$ is defined for positive times 
($0<\tau$) as follows:
\begin{eqnarray}    \label{e4_4}
\alpha (\tau) =
\int^{\infty}_0 \frac{d\omega}{\pi}
J(\omega) \sin(\omega\tau) =
-\frac{d}{d\tau} \int^{\infty}_0 \frac{d\omega}{\pi}
\frac{J(\omega)}{\omega} \cos(\omega\tau) \ \ \ \ .
\end{eqnarray}
 
In order to make further progress, a specification of the
initial state of the system plus the bath is needed.  We shall
assume that initially (at time $t=0$) the system is prepared
in some arbitrary quantal state while the bath oscillators
are in thermal canonical equilibrium with some reciprocal
temperature $\beta$. The Wigner function representation
of the probability density matrix is then 
(see App. A of Ref.\cite{ohmic}): 
\begin{eqnarray}  \label{e4_5}
\rho_{t=0} (R,P;Q_{\alpha},P_{\alpha}) = \rho_{t=0}(R,P)
\cdot \prod_{\alpha} \rho_{eq} (Q_{\alpha},P_{\alpha})    
\end{eqnarray}
where
\begin{eqnarray} \label{e4_6}
\rho_{eq}(Q_{\alpha},P_{\alpha}) = \
\frac{1}{\frac{1}{2} \coth (\frac{1}{2} \hbar\beta\omega_{\alpha})}
\cdot
\exp \left[ - \beta \left( \frac{\tanh(\frac{1}{2} \hbar\beta
\omega_{\alpha})}{(\frac{1}{2}\hbar\beta\omega_{\alpha})} \right)
\left( \frac{P^2_{\alpha}}{2m_{\alpha}} + \frac{1}{2} m_{\alpha}
\omega^2_{\alpha} Q^2_{\alpha} \right) \right] \ \ .        
\end{eqnarray}
Using (\ref{e4_6}) one obtains the expectation values 
\begin{eqnarray}
\langle \frac{P_{\alpha}(0)^2}{2m_{\alpha}}\rangle \ \ = \ \ 
\langle \frac{1}{2} m_{\alpha}\omega^2_{\alpha} Q_{\alpha}(0)^2 \rangle 
\ \ =  \ \ \frac{1}{4} \hbar\omega_{\alpha} 
\coth(\frac{1}{2} \hbar\beta\omega_{\alpha}) \ \ \ \ .
\nonumber 
\end{eqnarray}
Hence it is easily found that $\langle {\cal F}_{noise}(t)\rangle = 0$ 
while
\begin{eqnarray}
\langle {\cal F}_{noise}(t) {\cal F}_{noise^{\prime}}(t^{\prime})\rangle 
\ = \ \ - \sum_{\alpha} c_{\alpha}^2 
\langle Q_{\alpha}(0)^2 \rangle \cos(\omega_{\alpha}(t{-}t')) 
u_{\alpha}^{\prime}(x) u_{\alpha}^{\prime}(x') \ \ \ \ . 
\nonumber
\end{eqnarray}
Thus, one identifies that the noise term is characterized by 
the autocorrelation function (\ref{e2_4}) with
\begin{eqnarray}  \label{e4_7}
\phi (t-t^{\prime}) = \int^{\infty}_0 \frac{d\omega}{\pi}
J(\omega) \ \hbar\coth(\frac{1}{2} \hbar\beta\omega) 
\ \cos[\omega (t{-}t^{\prime})] 
\end{eqnarray}
The Langevin equation (\ref{e4_1}) together with 
(\ref{e4_2})-(\ref{e4_4}) and (\ref{e4_7}) 
constitutes an exact and complete description of the reduced dynamical
behavior of the system, as long as the system is considered 
to be classical in nature.

\subsection{Langevin Equation in the Absence of Disorder}

For the BM model $u_{\alpha}(x)=x$. The derivation of Langevin 
equation (\ref{e2_1}) leads to ${\cal F}_{noise}$ that satisfies 
(\ref{e2_2}). The latter may be interpreted as a special case 
of (\ref{e2_3}), provided equation (\ref{e2_5}) for $w(r)$ is used.  
However, in the expression for the retarded force 
(\ref{e4_3}), $w'(x(t){-}x(t'))$ is replaced by $x(t')$, rather than by 
$-(x(t){-}x(t'))$. The physical significance of this minor difference 
is discussed below. 
In the equation for the retarded force (\ref{e4_3}) one
may extend the integration over $t'$ to infinity, provided 
$\alpha(\tau)$ is replaced by 
$\tilde{\alpha}(\tau)=\alpha(\tau)\Theta(\tau)$, where 
$\Theta(\tau)$ is the step function. In turn, this kernel 
may be written as a sum of three terms, namely 
$\tilde{\alpha}(\tau)=\alpha_0(\tau)+\alpha(\tau)+\alpha_m(\tau)$ .
The kernel $\alpha(\tau)$ is the asymmetric continuation of 
$\tilde{\alpha}(\tau)$ to the domain $\tau<0$, 
while $\alpha_0(\tau)+\alpha_m(\tau)$ is the symmetric continuation.
The latter is split into $\alpha_0(\tau)$ which is a delta function,
and $\alpha_m(\tau)$ which satisfies 
$\int_{-\infty}^{\infty}\alpha_m(\tau)d\tau=0$.
Consequently, the retarded force, for the BM model, may be written 
as the sum:
\begin{eqnarray}   \label{e4_8}
{\cal F}_{retarded} \ = \ {\cal F}_{switching} + 
{\cal F}_{\Delta potential} + {\cal F}_{friction} +
{\cal F}_{\Delta mass}
\end{eqnarray}
where ${\cal F}_{switching}\approx -k_0\cdot\delta(t)\cdot x(t)$,
and ${\cal F}_{\Delta potential}\approx +\Delta k\cdot x(t)$,
and ${\cal F}_{friction}\approx -\eta_{eff}\cdot\dot{x}(t)$, 
and ${\cal F}_{\Delta mass}\approx -\Delta m\cdot\ddot{x}(t)$, 
with $k_0=\eta_{eff}=\lim_{\omega\rightarrow 0}\frac{J(\omega)}{\omega}$,
and $\Delta k = \frac{2}{\pi}\int_0^{\infty}\frac{J(\omega)}{\omega}d\omega$,
and $\Delta m = - \int_0^{\infty}\alpha(\tau) \ \tau^2 \ d\tau$.
It has been assumed that $\alpha(\tau)$ has short
range duration $\tau_c$, much shorter than the physically 
relevant time scales of the dynamics. {\em This assumption 
is not true in general}, and the consequences will be 
discussed later in the next section. 

The switching impulse act on the particle if it starts its 
trajectory at a point $x\ne0$. This term originates due to 
the fact that the initial preparation is such that the bath
is in thermal equilibrium provided $x=0$.  The $\Delta potential$
force may be avoided if we care to include in the BM Hamiltonian
the proper ``counter term''. Namely:
\begin{eqnarray}  \label{e_counter}
{\cal H}_{bath}+{\cal H}_{int} \ \ \ \rightarrow \ \ \ 
\sum_{\alpha}\left(\frac{P_{\alpha}^2}{2m_{\alpha}}
+\frac{1}{2} m \omega_{\alpha}^2    (Q_{\alpha}-
\frac{c_{\alpha}}{m_{\alpha}\omega_{\alpha}^2}x)^2 \right) 
\end{eqnarray}
This expression is manifestly invariant under space
translations. Sanchez-Canizares and Sols have further 
considered this issue \cite{sols}.

\subsection{Langevin Equation for Disordered Environment}

We turn back to the DLD model with ${\cal F}_{retarded}$
as found in equation (\ref{e4_3}). Here, performing similar treatment, 
neither ${\cal F}_{switching}$ nor
${\cal F}_{\Delta potential}$ are encountered. Formally, this
is due to the fact that the difference $-(x(t){-}x(t'))$ 
appears in (\ref{e4_3}), rather than $x(t')$ by itself. Physically,
this is due to the inherent homogeneity of the 
interaction with the environment (\ref{e3_2}).  As for 
${\cal F}_{friction}$ and ${\cal F}_{\Delta mass}$, here
one obtains
\begin{eqnarray}  \label{e4_11}
\ & \ & {\cal F}_{friction}(v) \ = \
\int_{-\infty}^{\infty}\alpha(\tau) \ 
w'(v\tau) \ d\tau  \nonumber \\
\ & \ & \Delta m \ = \ \int_0^{\infty}\alpha(\tau) \
w''(v\tau) \ \tau^2 \ d\tau
\end{eqnarray}
If $\alpha(\tau)$ has short range duration $\tau_c$, 
and if furthermore the velocity of the particle
is not too high ($v \cdot \tau_c \ll \ell$), then,  using 
$w'(0)=0$ and \mbox{$w''(0)=-1$}, the above expressions
reduce to those that follow equation (\ref{e4_8}).  
If $\alpha(\tau)$ is not short range, to be discussed below,
then the retarded force in the BM model is characterized by 
long range memory for the dynamics. The approximation that has 
been used in the preceding subsection is no longer valid, and 
infinities are encountered in the computations of the constants
there. In general, these annoying features are not shared by the 
DLD model ((\ref{e4_11}) with (\ref{e2_6})), since a finite cutoff 
$\tau_{\ell} = \ell/v$ exists, where the microscopic length 
scale $\ell$ characterizes the interaction range with 
the scatterers.

 
\section{Ohmic and Non-Ohmic Baths}
\setcounter{equation}{0}

In order to make further progress, a specification of 
the spectral function $J(\omega)$ is required. This 
function, defined in (\ref{e3_5}), characterizes the 
distribution of the bath-oscillators with respect
to their frequencies. Following CL we assume it to
be of the form 
\begin{eqnarray}   \label{e5_1}
J(\omega) \ = \ \eta\omega^s \ G(\omega/\omega_c). 
\ & \ & \mbox{for ``subohmic'' bath $0<s<1$} \nonumber \\
\ & \ & \mbox{for ``ohmic'' bath $s=1$} \nonumber \\
\ & \ & \mbox{for ``superohmic'' bath $1<s$} 
\end{eqnarray}
Here the exponent $s$ characterize the singular behavior
of $J(\omega)$ in the vicinity of $\omega=0$, 
while $G$ denotes a smooth cutoff function.
The latter satisfies $G(0)=1$. Moreover, it
is assumed that $G(|\omega|)$ is analytic for 
real frequencies. For example, $G$ may be chosen
to be either Lorentzian or Gaussian. The latter 
possibility will be adopted from now on. The commonly 
used Exponential cutoff $exp(-\omega/\omega_c)$
will not be used since it does not satisfies the
mentioned requirements. Exponential cutoff results in
singular behavior that corresponds to the
exponents $s,(s+1),(s+2),...$ rather than ``pure''
singular behavior that corresponds to $s$ alone. 
The asymptotic behavior of both $\alpha(\tau)$
and $\phi(\tau)$ is dictated by the singularity 
of $J(\omega)$ at $\omega=0$.  The physical significance 
of $s$ is further discussed in App.A and in Sec.VI(B).

\subsection{Expressions for the Kernels}

\begin{figure} 
\begin{center}
\leavevmode 
\epsfysize=6cm
\epsffile{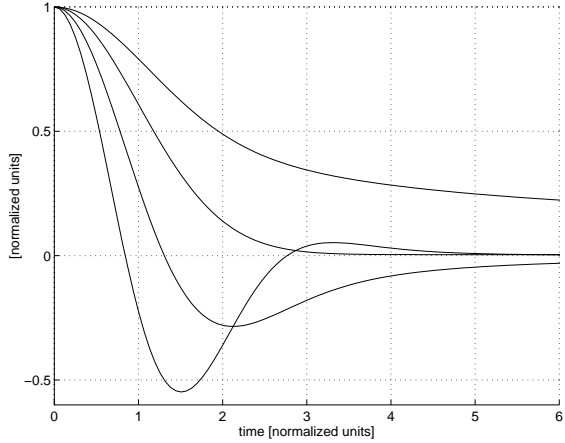}
\end{center}
\caption{\protect\footnotesize Plots of $G_s(\tau)$ 
 for $s=0.5$, $s=1$, $s=2$ and $s=4$. The plots are
 easily distinguished by referring to their 
 description in the text. }
\label{g_s_plot}
\end{figure}

In order to display explicit expressions for
both $\alpha(\tau)$ and $\phi(\tau)$, it is useful
to define the following function:
\begin{eqnarray}  \label{e5_4}
G_s(\tau) \ \equiv \ \int_0^{\infty} \frac{d\omega}{\pi} 
\ \omega^{s-1} \
\mbox{e}^{-\frac{1}{2}(\frac{\omega}{\omega_c})^2}
\ \cos(\omega\tau)
\end{eqnarray}
For $s=1$ this function is a normalized Gaussian. For
odd $s$ one obtains
\begin{eqnarray}   \label{e5_5}
G_{1{+}2n}(\tau) \ = \ (-1)^n \cdot \frac{d^{2n}}{d\tau^{2n}}
\left[\frac{\omega_c}{\sqrt{2\pi}}
\mbox{e}^{-\frac{1}{2}(\omega_c\tau)^2}\right]
\ \ \ \ \ \mbox{[n is integer]}
\end{eqnarray}
For general $s$ it starts at $\tau=0$ positive
\begin{eqnarray}  \label{e5_6}
G_s(0) \ = \ \frac{1}{2\pi} \ 2^{s/2}
\ \Gamma(\frac{s}{2}) \ \omega_c^{s}
\end{eqnarray}
and cross the $zero$ value $[(s+1)/2]$ times (now we
consider $s \ne$ odd number, possibly non-integer).
The short range oscillatory behavior dies out on a time scale
of the order $\tau_c=\omega_c^{-1}$. For larger times a power-law 
decay is found:
\begin{eqnarray}   \label{e5_7}
G_s(\tau)=\left(\frac{1}{\pi}\cos(\frac{\pi}{2}s)\Gamma(s)\right)
\cdot \frac{1}{\tau^s} \ \ \ \ \ \ \mbox{[tail for $s\ne$ odd]}
\end{eqnarray} 
The total algebraic ``area'' under $G_s(\tau)$ is infinite
for $0<s<1$, finite for $s=1$, and zero for $1<s$. 
Representative plots of $G_s(\tau)$ are displayed
in Fig.\ref{g_s_plot}.

For the spectral function $J(\omega)$ as in (\ref{e5_1}), with
Gaussian cutoff, the kernel $\alpha(\tau)$ is
\begin{eqnarray}   \label{e5_8}
\alpha(\tau) \ = \ -\eta \ \frac{d}{d\tau}G_s(\tau)
\end{eqnarray}  
Also $\phi(\tau)$ can be expressed in terms of $G_s(\tau)$
in both cases of ``high'' and ``low'' temperatures.  In the first 
case it is assumed that $\hbar\beta$ (that has dimensions of
time) is much shorter than any dynamical time scale. One
may use then the approximation 
\begin{eqnarray}   \label{e5_9}
\phi(\tau) \ = \ 2\frac{\eta}{\beta} \ G_s(\tau)
\ \ \ \ \ \ \ \mbox{[``high'' temperatures]}.
\end{eqnarray}
Else, if $\hbar\beta$ is long, one obtains
\begin{eqnarray}   \label{e5_10}
\phi(\tau) \ = \ 
\frac{2}{\pi}\Gamma(s{+}1)\zeta(s{+}1)
\cdot\frac{\eta}{\hbar^s}\left(\frac{1}{\beta}\right)^{s{+}1} 
\ + \  \hbar\eta G_{s{+}1}(\tau)
\ \ \ \ \mbox{[``low'' temperatures]},
\end{eqnarray}
where $\zeta(s)$ is Riemann Zeta function (sum over $1/n^s$). 
The notion ``low temperatures'' means here  $\tau<\hbar\beta$.
The temperature-dependent constant term results from the integration 
$\int_0^{\infty}\frac{d\omega}{\pi}J(\omega)\hbar
[\coth(\frac{1}{2}\hbar\beta\omega){-}1]$. 
From now on we shall use the notions of ``high'' and ``low'' 
temperatures in the sense of the above approximations.


\subsection{Expressions for Friction and Effective Mass}

We turn back to the computation of the terms in (\ref{e4_8}).
For the $\Delta potential$ term the 
result is always finite
\begin{eqnarray}  \label{e5_11}
\Delta k \ = \ \frac{1}{\pi} 2^{s/2} \ \Gamma(\frac{s}{2}) 
\ \eta \omega_c^s
\ \ \ \ \mbox{[switching impulse in BM model]} \ \ .
\end{eqnarray}
For the $friction$ one obtains a finite, non-zero result, 
($\eta_{eff}=\eta$) only in the case of an ohmic bath ($s=1$). 
For subohmic bath ($0<s<1$) the friction is {\em infinite}, while 
for superohmic bath ($1<s$) the friction is {\em zero}.
In case of the DLD model finite results are obtained
for all cases. Using (\ref{e4_11}) one finds out 
\begin{eqnarray}  \label{e5_12}
{\cal F}_{friction}(v) \ = \ \eta v \int_{-\infty}^{\infty} d\tau 
\int_0^{\infty} \frac{d\omega}{\pi} \  
\omega^s \ \mbox{e}^{-\frac{1}{2}(\frac{\omega}{\omega_c})^2}  
\ \sin(\omega\tau) \cdot \frac{d}{d\tau} \left[ 
\tau_{\ell}^2 \ \mbox{e}^{-\frac{1}{2}(\frac{\tau}{\tau_{\ell}})^2}
\right] \ \nonumber \\
\ \ = \ -\frac{2}{\pi}\Gamma(1{+}\frac{s}{2}) \
2^{s/2} \cdot \frac{\tau_{\ell}^3}
{ (\tau_c^2+\tau_{\ell}^2)^{1+\frac{s}{2}} }\cdot \eta v \
\ \rightarrow \ \
-\frac{2}{\pi}\Gamma(1{+}\frac{s}{2}) \
2^{s/2} \cdot \ell^{1-s} \cdot \eta v^s \
\end{eqnarray}
Above, the notation $\tau_{\ell}=\ell/v$ has been used. In
the last step the limit $\omega_c\rightarrow\infty$
has been taken since it leads to a finite, non-zero result.
Note that the cutoff frequency $\omega_c$ is not relevant 
physically as long as $v\ll\ell\cdot\omega_c$.  
The above computation clarifies the significance of 
the various time scales. Formally, the BM model
constitutes a special case with $\tau_{\ell}=\infty$.
For $s=1$, the so called ``ohmic'' case, one obtains 
${\cal F}_{friction}=-\eta v$. This result holds in the 
case of the DLD model as well as in the case of 
the BM model.  
In the latter case, for $s<1$ or for $1<s$, the friction 
force is either infinite or zero respectively. 
This is due to a long range memory effect. The cutoff 
frequency $\omega_c$, by itself, does not prevent this 
feature. Finite results for the components of retarded force 
in the BM model (\ref{e4_8}) may be obtained only for bounded 
systems, where $x(t)$ explores only a finite portion of the 
space. The calculation above (\ref{e5_12}) 
manifests the fact that a finite result, in case of
unbounded system, may be obtained if a physical cutoff 
$\tau_{\ell}$ is introduced. Such arise naturally  
in case of the DLD model.

Similar picture emerges upon calculation of the 
effective mass. For the BM model, a finite result
is obtained in case of an ohmic bath ($s=1$). 
\begin{eqnarray}    \label{e5_13}
\Delta m \ = \ \int_0^{\infty} \eta\frac{d}{d\tau}
\left[\frac{\omega_c}{\sqrt{2\pi}}
\mbox{e}^{-\frac{1}{2}(\omega_c\tau)^2} \right]
\cdot \tau^2 \cdot d\tau \ = \
-\sqrt{\frac{2}{\pi}}\ \frac{\eta}{\omega_c}
\ \ \ \ \ \ \ \ \ \ \ \ \ \mbox{for $s=1$}
\end{eqnarray}
Also for $2<s$, the power law tail of $\alpha(\tau)$
decays sufficiently fast to guarantee a finite result:
\begin{eqnarray}   \label{e5_14}
\Delta m \ = \ \frac{1}{2\pi} \  2^{s/2} \ 
\Gamma(\frac{s}{2}-1) \cdot \eta\omega_c^{s-2}
\ \ \ \ \ \ \ \ \ \ \ \ \ \mbox{for $2<s$} \ \ \ \ .
\end{eqnarray}
For $0<s<2$ ($s\ne 1$), the effective mass is 
infinite. In order to obtain a finite result
we should turn to the DLD model (\ref{e4_11}).
Here, as in the case of friction calculation, 
the spectral cutoff $\omega_c$ is not important,
and will be taken to infinity. Thus the integration
may be performed using the asymptotic expression
for $\alpha(\tau)$. Substitution of (\ref{e2_6}),
(\ref{e5_8}) and (\ref{e5_7}) yields
\begin{eqnarray}  \label{e5_15}
\Delta m \ \ & = & \ \ \frac{\eta}{\pi}\cos(\frac{\pi}{2}s)
\Gamma(s)\int_0^{\infty}\frac{-s}{\tau^{s{+}1}}
\frac{d^2}{d\tau^2}\left[\tau_{\ell}^2
\ \mbox{e}^{-\frac{1}{2}(\frac{\tau}{\tau_{\ell}})^2}
\right] \ \tau^2 \ d\tau \nonumber \\
\ \ & = & -\frac{s(s-1)}{\pi}\cos(\frac{\pi}{2}s)
\Gamma(s)\Gamma(1{-}\frac{s}{2}) \ 2^{-s/2} 
\cdot \eta\tau_{\ell}^{2-s}
\ \sim \ -\frac{\eta}{v^{s{-}2}}
\ \ \ \ \mbox{[for $0{<}s{<}2$]}
\end{eqnarray}
For the special case $s=1$ one obtains $\Delta m=0$
which is consistent with (\ref{e5_13}) in the limit 
$\omega_c\rightarrow\infty$. For $s\rightarrow 2$
a negative infinite value is obtained, while from
(\ref{e5_14}) a positive infinite value is obtained. In
order to describe correctly the crossover at 
$s=2$, one should introduce finite cutoff $\omega_c$
as well as finite $\tau_{\ell}$. The calculation will
not be carried here.


\section{Propagator For Non Dynamical Environment} 
\setcounter{equation}{0}

In this section we shall develop a path-integral expression 
for the propagator of a particle that interacts either with
static (quenched) or noisy environment. First we develop a classical
expression, and then we generalize to the quantal regime.

\subsection{Classical Derivation}
  
We refer to the dynamics generated by the Hamiltonian (\ref{e3_1}).
The classical Liouville propagator, over infinitesimal time $d\tau$,
and for definite realization of ${\cal U}$ is:
\begin{eqnarray}
{\cal K}(R_2,P_2|R_1,P_1)= 2\pi
\delta\left((P_2{-}P_1)+(\frac{\partial V}{\partial R}
{+}\frac{\partial {\cal U}}{\partial R})d\tau \right) \cdot
\delta\left((R_2{-}R_1)-\frac{P}{m}d\tau\right)
\nonumber
\end{eqnarray}   
It is more convenient to write an expression for the
propagator of the Fourier transformed probability function
(\mbox{$\rho(R,P)\rightarrow\rho(R,r)$}), namely,
\begin{eqnarray}
{\cal K}(R_2,r_2|R_1,r_1)=
(\frac{m}{2\pi\hbar d\tau}) \cdot
\exp\left[\frac{1}{\hbar}\left(
im(\frac{R_2{-}R_1}{d\tau})(r_2{-}r_1) \ - \
i(\frac{\partial V}{\partial R}{+}
\frac{\partial {\cal U}}{\partial R})
(\frac{r_2{+}r_1}{2})d\tau\right)\right]
\nonumber
\end{eqnarray}
Here a dummy parameter $\hbar$ has been inserted. Its
value does not have any effect here. However, later comparison
to the quantum mechanical version will be more transparent.
For finite time, the convolved propagator may be
written as a functional integral:
\begin{eqnarray}
{\cal K}(R,r|R_0,r_0)=
\int_{R(0)}^{R} \int_{r(0)}^{r} {\cal D}R {\cal D}r
\ \exp\left[\frac{1}{\hbar}\left(
im \int_0^t d\tau \dot{R}\dot{r} \ - \
i\int_0^t d\tau \frac{\partial V}{\partial R} \cdot r(\tau)
\right.\right. 
- \left. \left. i\int_0^t d\tau \frac{\partial {\cal U}}{\partial R}
\cdot r(\tau) \right) \right] \nonumber
\end{eqnarray}
Here the measure is
\begin{eqnarray} \label{e6_1}
{\cal D}R \ {\cal D}r \ = \ \ 
{ \left( \frac{m}{2\pi\hbar d\tau} \right) }
^{\mbox{\# segments}}
\ ...dR_3 dR_2 dR_1 \ ...dr_3 dr_2 dr_1
\end{eqnarray}
and the restrictions at the endpoints are $R(0)\!=\!R_0$, 
$r(0)\!=\!r_0$ and $R(t)\!=\!R$, $r(t)\!=\!r$. 
It is now possible to average over realizations of ${\cal U}$,
using the well known Gaussian identity
\begin{eqnarray} \label{e6_2}
\langle\mbox{e}^{i\int d\tau k(\tau)z(\tau)}\rangle_z \ = \ \ 
e^{-\frac{1}{2}\int d\tau d\tau'
\langle z(\tau)z(\tau')\rangle k(\tau)k(\tau')}
\end{eqnarray}
One obtains
\begin{eqnarray}  \label{e6_3}
{\cal K}(R,r|R_0,r_0)=
\int_{R(0)=R_0}^{R} \int_{r(0)=r_0}^{r} {\cal D}R {\cal D}r
\ \ \mbox{e}^{i\frac{1}{\hbar}S_{eff}[R,r]} 
\ \ \mbox{e}^{-\frac{1}{\hbar^2}S_N[R,r]}
\end{eqnarray}
where
\begin{eqnarray}  \label{e6_4}
S_{eff}[R,r]= S_{free}[R,r] -
\int_0^t d\tau \frac{\partial V}{\partial R} \cdot r(\tau) 
\ \ \ \ \ \ \mbox{[classical]}
\end{eqnarray}
and
\begin{eqnarray}  \label{e6_5}
S_N[R,r]=\frac{1}{2}\int_0^t\int_0^t d\tau_1 d\tau_2 \
[-\phi''(R(\tau_2){-}R(\tau_1),\tau_2{-}\tau_1)] \ r(\tau_2)r(\tau_1)
\end{eqnarray}
The free part of the effective action, for use in (\ref{e6_3}), is
$S_{free}[R,r]=m \int_0^t d\tau\dot{R}\dot{r}$.
It may be more convenient to write down the path-integral
expression for ${\cal K}(R,P|R_0,P_0)$. This expression is
obtained by double Fourier transform of ${\cal K}(R,r|R_0,r_0)$.
The result is
\begin{eqnarray}  \label{e6_7}
{\cal K}(R,P|R_0,P_0)=
\int_{R_0,P_0}^{R,P} \!\!\! {\cal D}R \int {\cal D}r 
\ \ \mbox{e}^{i\frac{1}{\hbar}S_{eff}[R,r]} 
\ \ \mbox{e}^{-\frac{1}{\hbar^2}S_N[R,r]}
\end{eqnarray}
Here, 
\begin{eqnarray}  \label{e6_8}
S_{free}[R,r]=-m \int_0^t d\tau\ddot{R}{r}
\ \ \ \ \ \ \ \mbox{[For use in (\ref{e6_7})]} \ \ \ .
\end{eqnarray}
Note that the integration ${\cal D}r$ is not restricted
at the end-points, whereas the integration ${\cal D}R$
is restricted at the end-points both in $R$ and in $\dot{R}$.
The restriction on $\dot{R}$ at the endpoints is implicit,
through the dispersion relation $\dot{R}=P/m$.

\subsection{Qauntal Derivation} 

A similar expression may be obtained for the quantal propagator.
Again, we refer to the dynamics generated by the Hamiltonian (\ref{e3_1}). 
The environment may be either ``noisy'' or ``quenched'', where the latter
case constitute formally a special case of the former. The expression
that will be obtained is a generalization of a result that  has been 
obtained in Ref.\cite{marianer} for white noise potential. 

The Feynman path-integral expression for the propagator
of the quantal wave-function is,
\begin{eqnarray}  \label{e6_9}
U(x|x_0)=\int_{x(0)=x_0}^{x} \!\!\!\! {\cal D}x 
\ \ \mbox{e}^{ i\frac{1}{\hbar}\int_0^t d\tau 
(\frac{1}{2}m\dot{x}^2 - V(x) - {\cal U}(x,t)) }  \ \ \ .
\end{eqnarray}
The path integral expression for the propagator
of the density probability function constitutes 
summation ${\cal D}x'{\cal D}x''$ over the pairs of paths
$x'(\tau)$ and $x''(\tau)$. Alternatively, we may use also 
the coordinates $R=(x'{+}x'')/2$ and $r=(x''{-}x')$, thus 
the summation will be ${\cal D}R{\cal D}r$, namely    
\begin{eqnarray}
{\cal K}(R,r|R_0,r_0)=
\int_{R_0}^{R} {\cal D}R \int_{r_0}^{r} {\cal D}r
\ \ \mbox{e}^{i\frac{1}{\hbar}S_{eff}[R,r] - 
i\frac{1}{\hbar}\int_0^t d\tau 
({\cal U}(x'',\tau)-{\cal U}(x',\tau))} 
\nonumber
\end{eqnarray}
where
\begin{eqnarray}   \label{e6_10}
S_{eff}[R,r]=S_{free}[R,r] -
\int_0^t d\tau (V(x'',\tau){-}V(x',\tau))  
\end{eqnarray}
It is important to notice that the quantal definition 
of the measure is identical with the classical one (\ref{e6_1}). 
In order to perform the average over realizations of ${\cal U}$
using the Gaussian identity (\ref{e6_2}), one may write the last 
expression as
\begin{eqnarray}
{\cal K}(R,r|R_0,r_0)=
\int_{R_0}^{R} \int_{r_0}^{r} {\cal D}R {\cal D}r
\ e^{i\frac{1}{\hbar}S_{eff}[R,r] 
-i\frac{1}{\hbar}\int_0^t d\tau \int_{-\infty}^{\infty} dz
(\delta(z{-}x'')-\delta(z{-}x')){\cal U}(z,\tau)} 
\nonumber
\end{eqnarray}
One easily find that the final result may be cast to the
form of equation (\ref{e6_3}) or (\ref{e6_7}) with
\begin{eqnarray}    \label{e6_11}
S_N[R,r]=\frac{1}{2}\int_0^t\int_0^t d\tau_1 d\tau_2 \ 
(\phi(x_2''{-}x_1'',\tau_2{-}\tau_1)
+\phi(x_2'{-}x_1',\tau_2{-}\tau_1)
-2\phi(x_2''{-}x_1',\tau_2{-}\tau_1))
\end{eqnarray} 
where $x_i$ is a short notation for $x(\tau_i)$. 

We are now in position to compare the classical propagator
((\ref{e6_7}) with (\ref{e6_4})-(\ref{e6_5})), with 
the quantal one ((\ref{e6_7}) with (\ref{e6_4})-(\ref{e6_5})). 
In the latter case $\hbar$ is, in
general, no longer a ``dummy variable''. The exception being the
case where the actions are quadratic in the path variables,
which is the case with BM model provided $V(x,t)$ is quadratic.
The ``classical feature'' may be characterized as arising from
invariance under the scaling transformation of the auxiliary
integration-variable $r(\tau)$. In the quantal regime the
replacement $\hbar\rightarrow\lambda\hbar$ cannot be compensated
by the scaling $r\rightarrow\lambda r$. Note however that the
limit $\hbar\rightarrow0$ is equivalent to taking leading 
behavior of the actions in the limit $r\rightarrow0$.


\section{Propagator for Dynamical Environment}
\setcounter{equation}{0}

\subsection{Feynman Vernon Formulation}

Here we follow closely the notations in Ref.\cite{CF}.  
The path-integral expression for the reduced propagator
of the probability density function is of the general 
form (\ref{e6_7}), with
\begin{eqnarray}   \label{e7_1}
S_{eff}[R,r]=S_{free}[R,r] -
\int_0^t d\tau (V(x'',\tau){-}V(x',\tau)) + S_F[x'',x'] 
\end{eqnarray}
The expressions for the reduced-action-functionals 
$S_F[x',x'']$ and $S_N[x',x'']$ in case of BM model, 
are given in equations (2.13)-(2.14) of the latter reference. 
In the BM model the interaction is via the dynamical
variable $x$, while, in DLD model, the
interaction is via $u_{\alpha}(x)$. Thus, in the
expressions for the friction functional $S_F[x',x'']$ 
and for the noise functional $S_N[x',x'']$  one should 
make the replacements \   
$x'  \rightarrow u_{\alpha}(x') \ , \ 
x'' \rightarrow u_{\alpha}(x'')$ \ ,
and sum over $\alpha$. Thus one obtains
\begin{eqnarray}
S_F[x',x''] \ = \ \frac{1}{\mbox{density}} \times  
\int_0^t\int_0^t d\tau_1 d\tau_2 
\ 2\tilde{\alpha}(\tau_2{-}\tau_1) \ \frac{1}{2} \sum_{\alpha}
(u_{\alpha}(x_2''){-}u_{\alpha}(x_2'))
(u_{\alpha}(x_1''){+}u_{\alpha}(x_1'))
\nonumber
\end{eqnarray}
and 
\begin{eqnarray}
S_N[x',x''] \ = \ \frac{1}{\mbox{density}} \times
\frac{1}{2} \int_0^t\int_0^t d\tau_1 d\tau_2 
\ \phi(\tau_2{-}\tau_1) \ \sum_{\alpha}
(u_{\alpha}(x_2''){-}u_{\alpha}(x_2'))
(u_{\alpha}(x_1''){-}u_{\alpha}(x_1'))
\ \ \ \ \ \ .\nonumber
\end{eqnarray}
Above $\tilde{\alpha}(\tau)\equiv\alpha(\tau)\Theta(\tau)$. 
These expressions may be simplified using (\ref{e3_6}).
The results are
\begin{eqnarray}  \label{e7_2}
S_F[x',x'']=\int_0^t\int_0^t d\tau_1 d\tau_2 
\ 2\tilde{\alpha}(\tau_2{-}\tau_1) \cdot \frac{1}{2} 
[w(x_2''{-}x_1'')-w(x_2'{-}x_1')+w(x_2''{-}x_1')-w(x_2'{-}x_1'')]
\end{eqnarray}
and
\begin{eqnarray}  \label{e7_3}
S_N[x',x'']=\frac{1}{2} \int_0^t\int_0^t d\tau_1 d\tau_2 
\ \phi(\tau_2{-}\tau_1) \ 
[w(x_2''{-}x_1'')+w(x_2'{-}x_1')-2w(x_2''{-}x_1')]
\end{eqnarray}
In the next paragraph we discuss further the physical significance 
of the functional $S_F[x',x'']$. 

For the BM model it has been noted before that in practice 
one could substitute (\ref{e2_5}). Thus
\begin{eqnarray}
S_F[x',x'']=\int_0^t\int_0^t d\tau_1 d\tau_2 
\ 2\tilde{\alpha}(\tau_2{-}\tau_1) \ r(\tau_2)\cdot (R(\tau_1)-R(\tau_2)) 
\ \ \ \ . \nonumber
\end{eqnarray}
However, the formally ``correct'' expression is somewhat 
different (Ref.\cite{CF} equation (2.13)). Namely, the 
integrand in the above equation is with $r(\tau_2)R(\tau_1)$, 
rather than $r(\tau_2)(R(\tau_1){-}R(\tau_2))$. One may say that
the BM reduced-action functional includes an additional term.
It is not difficult to demonstrate that the latter can be 
split into two terms that corresponds exactly to 
${\cal F}_{\Delta potential}$ and to ${\cal F}_{switching}$ 
discussed in Sec.III(B).  
The $\Delta potential$ term may be absorbed in the definition 
of $V(x)$, while the $switching$ term, which is 
$-\Delta k \ r(0)R(0)$, may be factored out of the 
path integral expression (see Ref.\cite{CF} equation (2.34)).    
It has the effect of operating on the initial probability 
function with
an impulse that acts on the particle if it starts its trajectory in
a point $x\neq0$. This term originates due to the fact that the 
initial preparation is such that the bath is in thermal equilibrium
provided $x=0$. Both the ``switching'' term and the additional 
effective potential, are absent in DLD model. This obvious results 
stems from the assumed inherent homogeneity of the environment.  
We turn now to the general expression (\ref{e7_2}). 
Again, as in Sec.III(B), it is convenient to express 
$\tilde{\alpha}(\tau)$ as a sum of its symmetric, 
and antisymmetric continuations. Here we focus on the 
resultant friction functional, which is
\begin{eqnarray}    \label{e7_5}
S_F |_{friction} \ = \ \ \int_0^t\int_0^t d\tau_1 d\tau_2 \ 
\alpha(\tau_2{-}\tau_1) \ w(x''(\tau_2){-}x'(\tau_1))
\end{eqnarray}
For Ohmic bath one obtains
\begin{eqnarray}   \label{e7_6}
S_F \ = \ \eta \int_0^t d\tau \ w'(r(\tau)) \ \dot{R}(\tau) 
\ \ \ \rightarrow \ \ \ -\eta \int_0^t d\tau \ r(\tau) \ \dot{R}(\tau) 
\end{eqnarray}
where in the last stage we indicated the classical limit.
Note that the classical expression for $S_F$ can be easily 
derived. For this one should include the friction force in 
the derivation of Sec.V(A).

\subsection{Derivation for Generic Bath}

A totally different derivation of the path integral expression 
for the propagator is possible in the general case of weak coupling 
to thermal, possibly chaotic bath.  This derivation, in  
case of {\em linear} coupling, has been introduce already 
by FV Ref.\cite{FV}.  The case of general, nonlinear coupling,
has been considered by M\"{o}hring and Smilansky Ref.\cite{uzy}.
Here we shall take a step further, and demonstrate that 
under ``normal'' circumstances, it will reduce to  an ohmic DLD
model.  We consider a bath Hamiltonian of the general form:
\begin{eqnarray} \label{e7_7}
{\cal H}_{bath}+{\cal H}_{int} \ = \
\sum_n |n\rangle E_n \langle n| \ + \ 
\sum_{nm} |m\rangle {\cal U}_{mn}(x) \langle n|  \ \ \ \ ,
\end{eqnarray} 
where $|n\rangle$ and $E_n$ are the eigenstates
and eigenenergies of the bath Hamiltonian in the
absence of the coupling.  This Hamiltonian depends
on $x$, the system variable, as a parameter. However, 
once the full Hamiltonian (\ref{e3_4}) is considered, 
$x$ becomes a dynamical variable. The so called  
{\em Influence Functional}, in our notations is 
(by definition):
\begin{eqnarray}  \label{e7_8}
\mbox{e}^{i(S_F+iS_N)} \ = \
\sum_{nm} p_n \ U_{mn}[x''(\tau)] \ U_{mn}^{\star}[x'(\tau)]
\ \ \ \ ,
\end{eqnarray}
Units with $\hbar=1$ are used here. $U[x(\tau)]$ is the 
evolution operator of the 
bath, in the presence of the ``driving force'' $x(\tau)$.  
The bath is assumed to be in canonical thermal equilibrium. 
The probability of the $n$-th eigenstate is 
$p_n\propto e^{-\beta E_n}$.
Using leading order perturbation theory one 
obtains
\begin{eqnarray} \label{e7_9}
U_{mn}[x(\tau)] \ \approx \ 
-i\int_0^t d\tau \ {\cal U}_{mn}(x(\tau)) \ 
\mbox{e}^{i(E_m-E_n)\tau} 
\ \ \ \ \ \ \mbox{for $m\ne n$} \ \ .
\end{eqnarray}
Similar expression holds for $m{=}n$ (see Ref.\cite{uzy}).
Substitution into (\ref{e7_8}) yields:
\begin{eqnarray} \label{e7_10}
(iS_F-S_N) \ \ = \ \ \sum_{nm} p_n \ \left[
\int_0^t \int_0^t d\tau_2 d\tau_1 \
{\cal U}_{mn}(x_2'') {\cal U}_{mn}^{\star}(x_1')
\ \mbox{e}^{i(E_m-E_n)(\tau_2-\tau_1)} \ \ \ \ \ \ \ \right.
\ \ \ \ \ \ \ \ \ \ \ \ \ \ \ \ \ \ \ \ \ \ \ \ \nonumber \\ 
\left.
-\int_0^t\int_0^{\tau_2} d\tau_2 d\tau_1  
\left(
{\cal U}_{mn}(x_2'') {\cal U}_{mn}(x_1'')
\ \mbox{e}^{-i(E_m{-}E_n)(\tau_2{-}\tau_1)} 
- 
{\cal U}_{mn}^{\star}(x_2') {\cal U}_{mn}^{\star}(x_1')
\ \mbox{e}^{i(E_m{-}E_n)(\tau_2{-}\tau_1)} 
\right) \right]  
\end{eqnarray}
Now we take a further assumption which will reduce 
the resultant expression for $S_F$ and $S_N$ to 
the form of (\ref{e7_2}) and (\ref{e7_3}) respectively.  
The matrix elements ${\cal U}_{mn}$ are assumed to 
be real, while their dependence on $x$ is assumed 
to be characterized by the function
\begin{eqnarray} \label{e7_11}
\pi \ \sum_{mn} \ (p_n-p_m) \ 
{\cal U}_{mn}(x_2) {\cal U}_{mn}(x_1)
\ \delta(\omega-(E_m{-}E_n)) \ \ = \ \
w(x_2{-}x_1) \cdot J(\omega) \ \ \ \ \ . 
\end{eqnarray}
The reduction of (\ref{e7_10}) to (\ref{e7_2}) and (\ref{e7_3}) 
is easily verified via algebraic manipulation, using 
$\Im[\sum_{nm}p_n\{..\}]{=}\sum_{nm}(p_n{-}p_m)\{..\}$
and 
$\Re[\sum_{nm}p_n\{..\}]{=}\sum_{nm}
\coth(\frac{1}{2}\beta\omega)(p_n{-}p_m)\{..\}$  
where 
$\coth(\frac{1}{2}\beta\omega){=}(p_n{+}p_m)/(p_n{-}p_m)$. 
Thus, for general nonlinear coupling, DLD model 
constitutes an equivalent representation for 
the bath, as far as the reduced dynamics of 
the system is concerned.  For the particular 
case of the weak {\em linear} coupling, there 
is further reduction to the BM model, with
\begin{eqnarray}  \label{e5_2}
J(\omega) \ = \ \pi\sum_{n,m} \ (p_n-p_m) \ |Q_{mn}|^2
\ \delta(\omega-(E_m{-}E_n))
\end{eqnarray}
In the above expression $Q$ denotes the collective bath
degree of freedom via which the interaction takes place,
namely, in (\ref{e7_7}) one should substitute 
${\cal U}(x)=Q\cdot x$. The bath variable $Q$ may be some 
complicated nonlinear combination of 
many elementary bath degrees of freedom. 
For both, the DLD model and the BM model, the expression
for $J(\omega)$ , may be cast into the form
\begin{eqnarray}  \label{e5_3}
J(\omega) \ = \ \ \frac{\pi}{\beta Z(\beta)} 
\ (1-\mbox{e}^{-\beta\omega})
\ \sigma(\omega)^2 \ K(\omega)
\end{eqnarray}
where $Z(\beta)$ is the partition function, 
$\sigma(\omega)$ is the standard 
deviation of the off-diagonal matrix elements
($\omega$ being the offset), and $K(\omega)$ is the 
density-density autocorrelation function of the 
spectrum $\{E_n\}$, appropriately averaged over 
the relevant energy scale.  The latter is related,  
for small energy differences, to the level spacing 
distribution.  However, for any practical use, 
one should ignore the effect of level spacing statistics 
on $J(\omega)$, since it corresponds to non-physically 
very long times.  Thus, the generic behavior of 
$J(\omega)$, for physically-relevant small $\omega$,
is expected to be $linear$.  This leads to the conclusion
that under ``normal'' circumstances the ohmic DLD model
is good representation for the dissipation process. 
This conclusion does not hold if {\em strong} coupling
to a chaotic bath is considered. In the latter case 
Zener transitions may dominate the dissipation 
process \cite{efrat,wilk}. We do not know whether the 
Influence Functional (\ref{e7_8}) for that case can be 
reduced to a form that resembles that of the DLD model.

\subsection{Loss of Quantal Interference}

The suppression of quantal interference is an important 
issue in both semiclassical and mesoscopic physics. 
It may arise that the quantum mechanical propagator 
may be expressed as a sum of probabilities to go 
either via one classical trajectory $x_a(\tau)$ or via
a different classical trajectory $x_b(\tau)$, plus an 
interference term.  The expression for the {\em Influence Functional} 
may be used in order to compute the suppression of 
the interference due to the interaction with the 
environment.  The interference term is multiplied 
by a ``dephasing'' factor $\langle e^{i\varphi} \rangle$, 
where we follow a notation due to Stern, Aharonov and Imry
\cite{stern}. In our notations, the dephasing factor is identified 
with $\exp(-S_N[x_a(\tau),x_b(\tau)])$. We defer further discussion 
of interference within the framework of FV formalism to
the last paragraph of this subsection.   

It is enlightening to consider the case of white noise, with 
$\phi(\tau_2{-}\tau_1)=\nu\delta(\tau_2{-}\tau_1)$.  For 
the BM model one obtains
\begin{eqnarray}  \label{e_supp}  
\langle \mbox{e}^{i\varphi} \rangle \  = \ \
\exp\left[-\frac{1}{2}\frac{\nu}{\hbar^2}\int_0^t 
\ (x_a(\tau)-x_b(\tau))^2 \ d\tau \right]  \ \ \ \ . 
\end{eqnarray}
Thus, interference is suppressed more effectively 
if the two interfering paths are better separated. 
A totally different result is obtained in case of the 
DLD model.  Here we assume that the two interfering 
paths are well separated with respect to the microscopic 
scale $\ell$, namely $\ell\ll|x_a(\tau){-}x_b(\tau)|$ most of 
the time.  It follows that 
\begin{eqnarray}  
\langle \mbox{e}^{i\varphi} \rangle \  = \ \
\exp\left[-\frac{\nu\ell^2}{\hbar^2} \cdot t \right]  \ \ \ \ . 
\end{eqnarray}
Here, the interference decays exponentially in time, 
and the actual spatial separation of the paths play
no role. Due to the disorder, dephasing events are as
effective for small separations as for large separations.

The dephasing of interference in metals due to 
electromagnetic fluctuations has been discussed by 
Al'tshuler, Aronov and Khmelnitskii in Ref.\cite{AAK}.
Their results have been re-derived by Stern, Aharonov
and Imry \cite{stern}. A somewhat simplified derivation 
is reconstructed in App.A. The strong dimensionality 
dependence of the dephasing process has been emphasized.  
This dependence is due to the separation-dependence as 
in (\ref{e_supp}), and see also (\ref{eA_supp1}).  
In case of the DLD model, the local nature of 
the dephasing process will eliminate this feature.

In order to understand how interference arise from the FV path
integral expression (\ref{e6_7}), it is convenient to rewrite 
it in the following form
\begin{eqnarray} \label{e_krp}  
{\cal K}(R,P|R_0,P_0) \ \ = \ \ 
\int_{R_0,P_0}^{R,P} {\cal D}R \ \ {\cal K}[R]  \ \ \ \ ,
\end{eqnarray}
where ${\cal K}[R]$ is a real functional, which is defined 
by the expression: 
\begin{eqnarray}  \label{e_kr}  
{\cal K}[R] =
\int_{unrestricted} \!\!\!\! {\cal D}r 
\ \ \mbox{e}^{i\frac{1}{\hbar}S_{eff}[R,r]} 
\ \ \mbox{e}^{-\frac{1}{\hbar^2}S_N[R,r]}
\ \ \ \ \ .
\end{eqnarray}
If the path integral expression for the evolution operator 
(\ref{e6_9}) is dominated by a {\em single} classical 
path $x_a(\tau)$, then the ${\cal D}R$ integration in 
(\ref{e_krp}) will be dominated by $R(\tau)=x_a(\tau)$.
Obviously, in order to obtain a non-vanishing result, the 
endpoint conditions should be compatible.  Turning to 
the computation of ${\cal K}[R(\tau){=}x_a(\tau)]$ via (\ref{e_kr}),
one observes that the ${\cal D}r$ integration is dominated 
by the trivial trajectory $r_R(\tau){=}0$.  We use the subscript $R$ 
in order to suggest that in general this trajectory 
should be $R$-dependent. Indeed, this is the case if two classical 
trajectories $x_a(\tau)$ and $x_b(\tau)$ dominate.  
One should consider then the ``interference path'' 
$R(\tau){=}(x_a(\tau){+}x_b(\tau))/2$, for which the ${\cal D}r$ 
integration is dominated by the non-trivial paths    
$r_R(\tau){=}\pm(x_a(\tau){-}x_b(\tau))$.  The existence of non-trivial 
path $r_R$ is the fingerprint of interference phenomena. A classical
trajectory $R(\tau)$ for which $r_R{=}0$ will not be damped, 
since $S_N[R,r_R]=0$ then. In contrast, an interference path, for 
which $r_R\ne 0$, is damped, since $0<S_N[R,r_R]$ in general. 
However, in Sec.IX, where localization effect is discussed,   
we shall encounter a vast family of interference trajectories
that are not damped by the noise functional.  In the latter 
case, the interference paths are found to be dominant in the 
computation of the propagator.   

Another issue that deserves attention is the interplay of 
friction and interference. By inspection of (\ref{e7_5}) it 
is clear that for {\em disordered} environment, quantal 
interference is unaffected by friction. This is true as long
as $x_a$ and $x_b$ are well separated in space. In Sec.VII(B) we
shall encounter a related quantal manifestation of this 
observation.  


\section{Spreading and Diffusion} 
\setcounter{equation}{0}

The main results of the two last sections are the path integral 
expression (\ref{e6_7}) for the propagator ${\cal K}(R,P|R_0,P_0)$,
with the appropriate action functionals $S_{free}$ (\ref{e6_8}), 
$S_F|_{friction}$ (\ref{e7_5}), and $S_N$ (\ref{e7_3}). The classical 
limit of $S_N$ is presented in (\ref{e6_5}). For ohmic friction, both 
the quantal version and its classical limit (\ref{e7_6}) will be 
further considered and compared. Friction in case of non-ohmic bath 
has been discussed in Sec.IV, and its quantal analog will not be 
considered further.

In order to get preliminary insight into the path-integral expression,
consider first the case of free particle in ``white'' noisy 
environment. Namely, $\phi(\tau{-}\tau')=\nu\delta(\tau{-}\tau')$,
with $\nu=2\eta k_BT$ as in \ref{e5_9}.  
In the classical case (\ref{e6_5}), one obtains 
\begin{eqnarray}      \label{e8_1}
S_N[r] = \frac{1}{2} \nu \int_0^t \ r(\tau)^2 d\tau \ \ \ .
\end{eqnarray}
independent of the spatial auto-correlation function $w(r)$. 
The observation that spatial correlations are of no importance, 
as long as the noise is uncorrelated in time, is trivial from 
classical point of view. In the quantum mechanical case, the 
corresponding expression is  
\begin{eqnarray}    \label{e8_2}
S_N[r] = \nu \int_0^t \ (w(0){-}w(r(\tau)) \ d\tau \ \ \ .
\end{eqnarray}
In contrary to classical intuition, spatial correlations may 
be of importance. However, for the BM model, (\ref{e2_5}), one 
recovers the classical result. 

In the path integral expression (\ref{e6_7}), one may perform the 
integration \mbox{${\cal D}r = ...dr_3dr_2dr_1$} \ . In the absence 
of $S_N$, each integration over $dr_n$ results in delta function
of velocities, namely $\delta(\dot{R}(\tau_{n+1})-\dot{R}(\tau_n))$.
In the presence of $S_N$, the integration $dr_n$ is weighted, and
as a result, each $\delta$ function is smeared. The propagator
constitutes a convolution of these smeared $\delta$ functions.
In particular, both in the classical case and in the BM model,
each smeared $\delta$ function is a Gaussian. It is obvious, that 
both in the classical and in the quantal case, $S_N$ leads to 
stochastic-like spreading. In what follow we want to estimate this 
spreading.


\subsection{Non Disordered Environment}

We turn now to estimate the spreading {\em in the absence of 
disorder}, which is the standard BM model. 
The noise functional is quadratic in the path-variable $r$, and is
independent of the path-variable $R$, namely 
\begin{eqnarray}   \label{e8_3}
S_N[R,r]=\frac{1}{2} \int_0^t\int_0^t d\tau_1 d\tau_2 
\ \phi(\tau_2{-}\tau_1) \ r(\tau)r(\tau')
\end{eqnarray}
Here, an exact treatment is available \cite{CF}. One may expand 
both $S_{free}$ and $S_N$ around the so-called classical paths, that 
are determined by the variation $\delta S_{free}=0$, with the
constraints $\delta R {=} \delta r {=} 0$ at the endpoints. 
The Gaussian integration 
is performed exactly. General expressions may be found 
in Ref.\cite{CF}. 
The phase space propagator ${\cal K}(R,P|R_0,P_0)$ is obtained
by double Fourier transform  of ${\cal K}(R,r|R_0,r_0)$, and obviously 
results in Gaussian function. In particular, one is interested in the 
spatial spreading. Setting $P_0=0$ and integrating over the final 
momentum $P$, one obtains
\begin{eqnarray}   \label{e8_5}
{\cal K}(R|R_0)= \frac{1}{ 2\pi\sqrt{\sigma_{spatial}} } 
\exp\left[\frac{1}{2}\left(
\frac{R{-}R_0}{\sigma_{spatial}}\right)^2\right]
\end{eqnarray}
A general expression for the spatial spreading, that applies
to any $\phi(\tau)$, may be obtained \cite{CF}: 
\begin{eqnarray}    \label{e8_7}
\sigma_{spatial}\ = \ \ \frac{1}{m} \ \sqrt{\int_0^t\int_0^t
\phi(\tau{-}\tau') \ r_{c\ell}(\tau) \ r_{c\ell}(\tau') \ 
d\tau d\tau' \ } \ \ \ ,
\end{eqnarray}
where $r_{c\ell}$ solves the linearized classical 
equation of motion ($m\ddot{r}+\eta\dot{r}=0$) 
with initial conditions $r(0)=0$ and $\dot{r}(0)=1$.
Note that ohmic friction is considered, for which the BM model is 
well defined. 
For the simplest case of white noise without friction one obtains 
$\sigma_{spatial} = \sqrt{\frac{1}{3}\nu\frac{t^3}{m^2}}$.
Friction leads to damping and diffusion. In the latter case, considering 
long time $e^{-\frac{\eta}{m} t} \ll 1$, one may disregard a short 
transient and substitute $r_{cl}=m/\eta$. Consequently 
$\sigma_{spatial}=\sqrt{\frac{\nu}{\eta^2}t}$. 
%
However, at low temperatures $\phi(\tau)=-\frac{C}{\pi}\frac{1}{\tau^2}$ 
while $\int_{-\infty}^{\infty}\phi(\tau)d\tau=0$. Thus
\begin{eqnarray}    \label{e8_9}
\sigma_{spatial} \ \sim \ \ \sqrt{\frac{C}{\eta^2} \cdot \frac{2}{\pi}\ln t}
\ \ \ \ \ \ \ \mbox{special case - suppressed diffusion} \ , 
\end{eqnarray}
diffusion is suppressed due to the negative autocorrelations
of the noise. This effect is classical in nature. Intentionally
we did not use the explicit expression for the constant, namely 
$C=\hbar\eta$.  The presence of $\hbar_{bath}$ in the 
formula, rather than $\hbar_{system}$, without explicit subscript,
may miss-lead the reader. The particle can be treated as a classical 
object, within the framework of e.g. Langevin equation, and still 
the suppression of diffusion will occur.


\subsection{Disordered Environment, White Noise} 

Both the classical limit and the BM model generate ``classical''
dissipation effects. We turn now to the DLD model, ((\ref{e8_2})
with (\ref{e2_6})). Here the situation is quite different. 
In order to compute the propagator, a different, more 
powerful strategy is required. We shall exploit the fact 
that for white noise $S_N[R,r]$ is still independent 
of $R(\tau)$, while $S_{eff}[R,r]$ is linear in $R(\tau)$.  
It is most convenient to Fourier transform ${\cal K}(R,r|R_0,r_0)$ 
to ${\cal K}(p,r|p_0,r_0)$.  The path integral expression may
be written for this representation as follows {\em (we suppress
from now on printing $\hbar$, but shall restore it later)}:
\begin{eqnarray}
{\cal K}(p,r|p_0,r_0)=\mbox{e}^{i\eta(Rw'(r){-}R_0w'(r_0))}
\int_{r_0,p_0}^{r,p}\!\!\!\!\!\! {\cal D}r \int {\cal D}R \
\mbox{e}^{-i\int_0^td\tau(m\ddot{r}+\eta w''(r)\dot{r})R}
\ \mbox{e}^{-\nu\int_0^t[w(0){-}w(r(\tau))]d\tau}
\nonumber
\end{eqnarray}
The ${\cal D}R$ integration can be performed now, yielding 
a delta function $\delta(m\ddot{r}+\eta w''(r)\dot{r})$ at 
each point along the trajectory. The trajectory for which 
the integrand does not vanish will be denoted by $r(r_0,r,\tau)$,
where $0<\tau<t$. This trajectory is required to satisfy  the  
endpoints conditions $r(0)\!=r_0$ and $r(t)\!=\!r$. Hence,
the result of the path-integration is 
\begin{eqnarray}
{\cal K}(p,r|p_0,r_0) \ = \ \ \mbox{e}^{i\eta(Rw'(r){-}R_0w'(r_0))}
\ \ \delta(p-m\dot{r}(r_0,r,t)) \ \ \delta(p_0-m\dot{r}(r_0,r,0))
\cdot \mbox{e}^{-\nu\int_0^t[w(0){-}w(r(r_0,r,\tau)]d\tau}
\nonumber
\end{eqnarray}
The Inverse Fourier transform yields
\begin{eqnarray}
{\cal K}(R,r|R_0,r_0) \ = \ \ \mbox{e}^{i\left[
(m\dot{r}(r_0,r,t)+\eta w'(r))R -
(m\dot{r}(r_0,r,0)+\eta w'(r_0))R_0 
\right] }
\ \ \mbox{e}^{-\nu\int_0^t[w(0){-}w(r(r_0,r,\tau)]d\tau}
\nonumber
\end{eqnarray}
To simplify the latter expression we note that 
$m\dot{r}{+}\eta w(r)$ is a constant of the motion.

The propagator ${\cal K}(R,P|R_0,P_0)$ is the Fourier 
transform in the variables $r$ and $r_0$. In order to 
get insight we restrict ourselves to the reduced
kernel ${\cal K}(R|R_0)$, namely
\begin{eqnarray}   \label{e9_1}
{\cal K}(R|R_0) \ = \ \ \int_{-\infty}^{\infty}dr_0
\ \ \mbox{e}^{i\left[
(m\dot{r}(r_0,0,0)+\eta w'(r_0))(R-R_0) 
\right] }
\ \ \mbox{e}^{-\nu\int_0^t(w(0){-}w(r(r_0,0,\tau))d\tau}
\end{eqnarray}
As before we distinguish the case of frictionless propagation,
for which $r(r_0,0,\tau)=(t{-}\tau)/t{\cdot}r_0$, from the case
of damped particle with $e^{-\frac{\eta}{m} t}\ll 1$. In the
latter case the trajectory is modified for $|r_0|{<}\ell$, where
$r(r_0,0,\tau)\approx r_0$. Consequently the ``phase'' in 
(\ref{e9_1}) is $S_{eff}(r_0){=}{-}m/t{\cdot}(R{-}R_0){\cdot}r_0$ and 
$S_{eff}(r_0){=}{-}\eta{\cdot}(R{-}R_0){\cdot}r_0$ for the 
two corresponding types of trajectories. As for the noise 
argument in (\ref{e9_1}), it is $S_N(r_0)=\nu\ell^2 t$ for 
$\ell\ll|r_0|$, and $S_N(r_0)\approx\frac{1}{6}\nu t r_0^2$ or
$S_N(r_0)\approx\frac{1}{2}\nu t r_0^2$ for $|r_0|{<}\ell$, depending 
whether friction is absent or present respectively. The separation
of scales both in $S_{eff}(r_0)$ and in $S_N(r_0)$ enables
splitting the integral in a convenient way, namely
\begin{eqnarray}
\int_{-\infty}^{\infty} ... \ dr_0 \ = \ \ 
\int_{-\infty}^{\infty} ... \ W\left( \frac{r_0}{\ell} \right) 
\ dr_0 \ \ + \ \ 
\int_{-\infty}^{\infty} ... \ \left(
1 - W\left(\frac{r_0}{\ell}\right) \right) \ dr_0   
\ \ \ \ ,
\nonumber
\end{eqnarray}
where $W(x)$ is a smooth, symmetric cutoff function that equals
$\approx 1$ for $|x|<1$ and equals $\approx 0$ for $1<|x|$.       
The integration in (\ref{e9_1}) is performed, and the following
expression is obtained for the propagator.
\begin{eqnarray}    \label{e9_2}
{\cal K}(R|R_0) \ \ \approx \ \ \ 
\tilde{W}\left(\frac{R}{\hbar/\eta\ell}\right)
\ \star \ {\cal K}_{c\ell}\left(R-R_{c\ell}(t)\right) 
\ \ + \ \ \mbox{e}^{-\frac{\nu\ell^2}{\hbar^2}t}
\ \cdot \ \left[\delta\left(R-R_0(t)\right)-
\tilde{W}\left(\frac{R-R_0(\tau)}{\hbar t/m\ell}\right) \right]
\ \ \ .
\end{eqnarray}
The symbol $\star$ stands for convolution, and results in smearing
of the classical propagator on scale $\hbar/(\eta\ell)$. 
For frictionless propagation one should use the 
replacement $\hbar/(\eta\ell)\rightarrow\hbar t/(m\ell)$. 
We have used the notation $R_0(t){=}R_{c\ell}(t){=}R_0$. The 
above result holds also if $P_0\ne 0$, in this case 
$R_0(t)$ will propagate as if friction is absent, while
$R_{c\ell}(t)$ will propagate as in the classical limit.
Obviously, if friction is indeed absent, then 
again $R_0(t)$ and $R_{c\ell}(t)$ will coincide.
The kernel ${\cal K}_{c\ell}(R|R_0)$ denotes the classical 
result, (\ref{e8_5}). 

The expression for the quantal propagator demonstrates
that a piece of the wavepacket is frozen due to the 
disorder. This is a non trivial quantal effect that indeed 
can be entitled ``Quantum Dissipation''. We emphasize again that 
such quantal effect is absent in the BM model.  However, 
the expression for the propagator also demonstrates that the 
``quantal correction'' goes to zero exponentially in time, 
as in the case of interference discussed in Sec.VI(C).


\section{Classical Non-Markovian Effects} 
\setcounter{equation}{0}

The disadvantage of the treatments that have been presented 
in the preceding section, is the difficulty to extend them to 
the {\em general} case of disordered environment, namely if the 
disorder (the noise) is correlated in time. We therefore
turn to a somewhat more heuristic approach, that will enable 
approximated treatment. 
For the computation of ${\cal K}(R,P|R_0,P_0)$ we shall use 
the classical limit (\ref{e6_5}) of $S_N[R,r]$. The 
``quantal correction'' in (\ref{e9_2}), is not considered 
again in the present section.  

In the classical limit (\ref{e_krp}) constitutes a formal 
solution of Langevin Equation.  The real functional 
${\cal K}[R]$ has a simple probabilistic interpretation.
For ohmic friction it takes the form 
\begin{eqnarray}   \label{e10_2}
{\cal K}[R] \ \ = \ \ \ \int {\cal D}r 
\ \ \mbox{e}^{-i\int_0^t d\tau \ \Delta_R(\tau) \ r(\tau)}
\ \ \mbox{e}^{-\frac{1}{2}\int_0^t\int_0^t d\tau d\tau'
\ \phi_R (\tau,\tau') \ r(\tau) \ r(\tau')}
\end{eqnarray}
where
\begin{eqnarray}   \label{e10_3}
\phi_R(\tau,\tau') \ \equiv \ \ - w''(R(\tau){-}R(\tau'))
\ \cdot \ \phi(\tau{-}\tau')
\end{eqnarray}
and
\begin{eqnarray}  \label{e10_4}
\Delta_R(\tau) \ \equiv \ \ m\ddot{R}(\tau) + \eta \dot{R}(\tau)
\end{eqnarray}
Formally, the unrestricted ${\cal D}r$ integration may
be performed exactly, yielding the result
\begin{eqnarray}   \label{e10_5}
{\cal K}[R] \ = 
\ \sqrt{\det[\Phi_R]}\cdot
\exp\left(-\frac{1}{2}\int_0^t\int_0^t
d\tau d\tau' \ \Phi_R(\tau,\tau') 
\ \Delta_R(\tau) \ \Delta_R(\tau')\right)
\end{eqnarray}
where $\Phi_R$ is the reciprocal of $\phi_R$. 
In order to compute ${\cal K}(R,P|R_0,P_0)$ one should identify
the most contributing paths, for which ${\cal K}[R]$ is maximal.
In subsection A, where we discuss short-time correlated colored noise, 
we assume that {\em one} optimal path dominates the computation.  
In subsection B, where we discuss static or almost-static 
noisy potential, we shall identify a whole family of optimal paths. 
In the next section, where quantum localization is discussed, 
we shall use a similar strategy, and a family of interference paths 
will be identified.


\subsection{Normal, Dissipative Diffusion}

In this section we shall analyze the diffusive behavior 
which is encountered in the absence of disorder. Our assumption 
will be that the ${\cal D}R$ integration is dominated by 
{\em one} smooth ``optimal path''.  We shall substantiate this 
assumption by demonstrating consistency with the exact result 
that has been presented in Sec.VII(A). Furthermore, it will be 
argued that the ``optimal path'' for short-time correlated noise
is the same as for white noise.    

\begin{figure} 
\begin{center}
\leavevmode 
\epsfysize=6cm
\epsffile{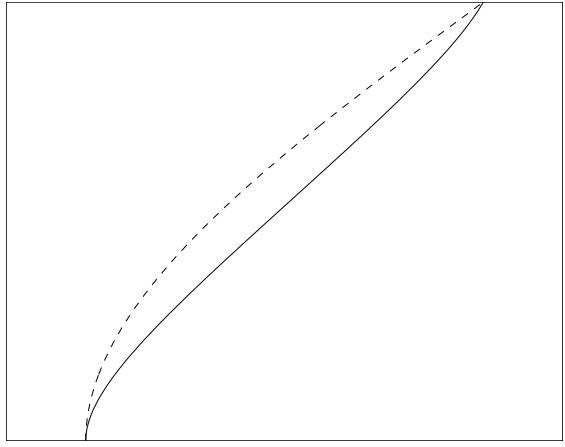}
\end{center}
\caption{\protect\footnotesize Illustration of the 
 optimal path (\protect\ref{e10_6}) for either 
 damped propagation (solid line), or frictionless 
 propagation (dashed line).  
 The horizontal axis represents spatial position, 
 while the vertical axis is for the time. }
\label{traj_dif}
\end{figure}

In order to find the path $R_o(\tau)$ that maximize 
${\cal K}[R]$, we consider first the case of white 
noise, where $\phi_R(\tau{-}\tau')=\nu_R\delta(\tau{-}\tau')$. 
Hence $\Phi_R(\tau{-}\tau')=\nu_R^{-1}\delta(\tau{-}\tau')$,
with $\nu_R=\nu$ (the subscript $R$ is reserved for later use).
As in Sec.VII(A) we focus the attention on the computation of the 
reduced propagator ${\cal K}(R|R_0)$. Formally, the path integral 
expression for ${\cal K}(R|R_0)$ is identical with (\ref{e_krp}), 
except for the restriction at the endpoints. For ${\cal K}(R|R_0)$, 
the relaxed constraints are $R(0){=}R_0$, \ $\dot{R}(0){=}0$, 
and $R(t){=}R$. Denoting $\dot{R}=v$, the variational equation for
$R_o(\tau)$, including Lagrange multiplier, is  
\begin{eqnarray} 
\delta \ \int_0^t [(m\dot{v}+\eta v)^2 + const\cdot v] d\tau = 0
\nonumber
\end{eqnarray}
one obtains
\begin{eqnarray}
& \ & m\ddot{v}-\eta^2 v = const \nonumber \\
& \ & (m\dot{v}+\eta v)_t = 0 \nonumber \\
& \ & v(0) = 0 \nonumber \\
& \ & \int_0^t v(\tau) d\tau = (R{-}R_0) \nonumber 
\end{eqnarray}
The last two equations are the constraints.
The solution for damped propagation   
($e^{-\frac{\eta}{m}t}\ll 1$) is easily found.  
For sake of comparison also the solution for 
frictionless propagation ($\eta=0$) is displayed.   
\begin{eqnarray}   \label{e10_6}
\ & \ & \dot{R}_o(\tau) \ \approx \ \ 
\frac{R{-}R_0}{t} \left( 
1-\frac{1}{2}\mbox{e}^{-\frac{\eta}{m} (t-\tau)}
-(1-\frac{1}{2}
\mbox{e}^{-\frac{\eta}{m} t})
\mbox{e}^{-\frac{\eta}{m}\tau}
\right) \ \ \ \ \mbox{for damped propagation} 
\nonumber \\
\ & \ & \dot{R}_o(\tau) \ = \ \ 
\frac{3}{2}\frac{R{-}R_0}{t^3} \tau (2t-\tau)
\ \ \ \ \mbox{for frictionless propagation} 
\end{eqnarray}
In the first formula a constant prefactor that 
equals $\approx 1$ has been dropped, since we
assume here $e^{-\frac{\eta}{m} t}\ll 1$ . The optimal paths 
of (\ref{e10_6}) are illustrated in Fig.\ref{traj_dif}.
Computation of $\Delta_R(\tau)$ for the optimal path is 
straight forward. For damped propagation the computation 
is trivial since $\Delta_R\approx\eta\frac{R{-}R_0}{t}$  
is constant. Substitution into (\ref{e10_5}) yields:
\begin{eqnarray}   \label{e10_7}
{\cal K}[R_o] \ = \ \ const\cdot\exp\left(-\frac{1}{2}\cdot
\frac{(R{-}R_0)^2}{\frac{1}{\eta^2}\ \nu_R \ t}\right) \ \ \ ,
\end{eqnarray}
which is in consistency with the exact result of Sec.VII(A).
It is easily verified that also for frictionless 
propagation, consistency with the exact result is maintained. 

For short-time correlated colored noise ($1\le s$) it is natural 
to replace $\phi(\tau{-}\tau')$ by $\nu\delta(\tau{-}\tau')$, 
with the effective white noise intensity $\nu=\phi(\omega{=}0)$.  
As long as $0<\nu$ (finite temperatures), the long time behavior 
is diffusive, and consistency with (\ref{e8_7}) is easily 
verified. Still, a more elaborated argument is required in order 
to substantiate the ``white noise approximation''. This argument 
will be discussed now. 

By inspection of (\ref{e10_5}) it is clear that the most 
contributing paths, for which ${\cal K}[R]$ is large, 
must satisfy $|\Delta_R(\omega)|^2<\phi(\omega)$. 
For white noise $\phi(\omega){=}\nu{=}const$, but still the 
``optimal path'' is smooth.  By ``smooth'' one means that
$\Delta_R(\omega)$ is concentrated within the interval 
$\omega<1/\tau_{\eta}$ where $\tau_{\eta}{=}m/\eta$ is the 
relevant time scale for the system's dynamics.  
Consider now a noise auto-correlation function of the form
$\phi(\tau)=\nu\delta(\tau)+C{\cdot}G_s(\tau)$.  
Its Fourier transform satisfies $\phi(\omega)\approx\nu\approx const$
for $\omega<(\nu/C)^{1/(s{-}1)}$.  Thus, the first requirement 
for the ``white noise approximation'' to hold should be 
$C/\tau_{\eta}^{s{-}1}<\nu$.  For $\omega_c<\omega$ the spectral  
function $\phi(\omega)$ drops to zero, which implies that such high
frequency components are not favored. So far there is consistency 
with our assumption that the most contributing paths are smooth. 
However, in the vicinity of $\omega_c$ the spectral function 
$\phi(\omega)$ is peaked. It implies, that unlike the case of 
white noise, an oscillatory component with time period $\tau_c$ is 
favored. Obviously, such component arise from the strong 
accelerations that the particle experiences within short 
periods whose duration is $\tau_c$.  Over these short periods 
the maximum displacement is 
$\Delta L=\frac{1}{2}(\sqrt{\phi(\tau{=}0)}/m)\tau_c^2$. 
Using (\ref{e5_6}) it is found that this amplitude is 
proportional to $\tau_c^{(4{-}s)/2}$.  For $s<4$ the amplitude $\Delta L$
goes to zero as $\tau_c\rightarrow0$. Therefore, in this 
restricted regime ($1\leq s<4$), and in particular for $s=2$ (low temperature 
ohmic noise) the ``white noise approximation'' should 
be adequate.


\subsection{Anomalous "Diffusion"}

Encouraged by the consistency of the heuristic
approach with exact results, we turn now to analyze 
the diffusion due to short-time correlated noise 
{\em in the presence of disorder}. We shall 
use the ``white noise approximation'' whose validity 
has been discussed in the preceding subsection. 
Namely, for short range correlated noise, the most 
contributing paths are concentrated around the same 
$R_o(\tau)$ that has been found for white noise. 
From (\ref{e10_6}) it follows that $R_o(\tau)$
is, up to end point transients, a free-like propagation.  
Consequently, the effective noise autocorrelation
function is
$\phi_R(\tau)\approx 
-w''((R{-}R_0)/t\cdot\tau)\cdot\phi(\tau)$
and we define
\begin{eqnarray}    \label{e10_8}
\nu_R \ \  = \ \ \ \int_{-\infty}^{\infty} \ \ 
[-w''\left(\frac{R{-}R_0}{t}\tau\right) \cdot \phi(\tau)]
\ d\tau
\end{eqnarray}
In general $0<\nu_R$, also in the limit of zero temperature.

In the presence of disorder, the effective white noise intensity 
is (in general) a function of the endpoint conditions, rather than 
a constant. For typical noise autocorrelation function of the 
form $\phi(\tau)=C\cdot G_s(\tau)$ one obtains
\begin{eqnarray}  \label{e10_9}
\ & \ & \nu_R \ =  \ \ \tilde{C} \cdot 
\frac{\tau_{\ell}^3}{(\tau_c^2+\tau_{\ell}^2)^{1+\frac{s}{2}}}
\nonumber \\
\ & \ & \mbox{with} \ \ \ \ \ \ \ 
\tau_{\ell} \ = \ \ \frac{\ell \cdot t}{(R{-}R_0)} 
\nonumber \\  
\ & \ & \mbox{and} \ \ \ \ \ \ \ 
\tilde{C} \ = \ \ \sqrt{\frac{2}{\pi}} \ 2^{s/2} 
\ \Gamma(1{+}\frac{s}{2}) \cdot C
\end{eqnarray}
The computation has been carried out by taking in 
(\ref{e10_8}) the Fourier transform of both $\phi()$ and $w''()$, 
and performing $d\omega$ integration rather than $d\tau$ 
integration. Substitution into (\ref{e10_7}) suggests that:
\begin{eqnarray}    \label{e10_10}
\ & \ &  
{\cal K}(R|R_0) \ |_{\{|R{-}R_0| < \ell\omega_c t\}} \ = \ 
const \cdot \exp\left( -\frac{\eta^2}{2\tilde{C}}
\ \ell^{s{-}1} \ t^{s{-}2} \ |R{-}R_0|^{3{-}s} \right)
\nonumber \\ 
\ & \ & 
{\cal K}(R|R_0) \ |_{\{|R{-}R_0| > \ell\omega_c t\}} \ = \ 
const \cdot \exp\left( -\frac{\eta^2}{2\tilde{C}}
\ \frac{|R{-}R_0|^5}{ \omega_c^{s{+}2} \ \ell^3 \ t^4 } \right)
\end{eqnarray}
The ``tail'' of the dispersion profile is universal.
It depends on $\omega_c$, but it is independent of 
the nature of the noise.
In contrast, the short range profile is determined
by the low frequency bath-oscillators, and thus it 
is sensitive to the exact value of $s$. For $s=1$,
(ohmic model, high temperatures), normal diffusive 
behavior prevails. For $s=2$ (ohmic model, low 
temperatures) the diffusion freezes. The dispersion
profile is Exponential rather than Gaussian, namely
\begin{eqnarray}    \label{e10_11}
{\cal K}(R|R_0) \ = \ \ 
const \cdot \exp\left( 
-\frac{|R-R_0|}{\left[
4\sqrt{\frac{2}{\pi}} \ \frac{1}{\eta^2 \ \ell} 
\ C \right]}   \right)
\ \ \ \ \ \ \ \mbox{[ohmic, low temperatures]}
\end{eqnarray}
The dispersion here, in the DLD model, is of the order 
$C/(\eta^2\ell)$ rather than $\sqrt{C}/\eta$. 
(BM model (\ref{e8_9})). One should observe that on 
for $C\sim(\eta\ell)^2$ both models predict 
consistently dispersion on spatial scale $\ell$. 
For larger noise intensity, the BM model is not valid, 
and DLD predicts always a larger dispersion, which is 
intuitively expected. For weak noise ($C<(\eta\ell)^2$)
the spatial spreading is on scale less than $\ell$. 
Our treatment of the DLD model is not valid on this 
microscopic scale. However, in this regime the BM model 
can be trusted. 
For $3<s<4$  equation (\ref{e10_10}) implies that the particle 
is evacuates from the vicinity of $R_0$.

The ``body'' of the ``diffusion'' profile (\ref{e10_10}a)
is determined by a low velocity paths for which
$\tau_c\ll\tau_{\ell}$. It is easily verified that 
a sufficient condition for the validity of the ``white
noise approximation'' is $\tau_{\ell}\ll\tau_{\eta}$.
This condition is satisfied in the relevant spatial range
($|R{-}R_0| < \ell\omega_c t$), except for a relatively 
small interval around $R_0$, which is determined by the
large ratio $\tau_{\eta}/\tau_c$.

The ``tail'' of the ``diffusion'' profile (\ref{e10_10}b)
is determined by a high velocity paths for which
$\tau_{\ell}\ll\tau_c$. Here the the validity argument 
should be modified. The spectral function $\phi(\omega)$
is peaked around $\omega=1/\tau_{\ell}$ rather than
around $\omega=\omega_c$.  Thus, oscillatory component 
which is characterized by period $\tau_{\ell}$ is favored.
The maximal spatial amplitude of this component is 
$\Delta L=\frac{1}{2}(\sqrt{\phi(\tau{=}0)}/m)\tau_{\ell}^2$. 
It is convenient to use a special notation for the standard
deviation of the disordered potential, namely  
$W=\ell\sqrt{\phi(\tau{=}0)}$.  Using this notation the
amplitude is $\Delta L=\frac{1}{2}(W/mv^2)\ell$.  This 
amplitude is required to be much less than $\ell$, or 
alternatively $W\ll\frac{1}{2}mv^2$ or alternatively
$\tau_{\ell}\ll\tau_W$, with $\tau_W=\ell/\sqrt{2W/m}$.  
In this subsection we have limited the discussion to the 
case of short-time correlated noise for which $\tau_c$ is 
small in some sense. Indeed, if it is assumed that 
$\tau_c\ll\tau_W$, then the validity condition will be 
satisfied automatically.


\subsection{Classical Localization and Non-Dissipative Diffusion} 
 
In this subsection we shall discuss the case of static or 
almost-static disordered non-dynamical environment ($\eta=0$). 
The ``noise'' is assumed to possess long-time correlations. 
Specifically:
\begin{eqnarray}    \label{e10_13}
\phi_R(\tau,\tau') \ = \ - w''(R(\tau){-}R(\tau'))
\ \cdot \ \frac{W^2}{\ell^2}
\exp\left(-\frac{1}{2}\frac{(\tau{-}\tau')}{\tau_c}^2\right)
\ \ \ \ ,
\end{eqnarray}
where $W$ is the standard deviation of the disordered potential.
Here $\tau_c$ is assumed to be large, much larger than
$\tau_W=\ell/\sqrt{2W/m}$. 

In general the ``white noise approximation'' breaks 
down. The path integral expression is no-longer dominated 
by ``smooth'' trajectories.  It is difficult to use 
the explicit formula (\ref{e10_5}) in order to identify
the family of ``optimal paths'' since $\Phi(\tau,\tau')$
is no longer diagonal. We therefore prefer to use 
heuristic considerations. No new insight is gained if one 
insists using (\ref{e10_5}).      
  
We take first the limit $\tau_c\rightarrow\infty$. The classical 
mean free path of the particle is 
\begin{eqnarray}    \label{e10_14}
L_{collision}\ = \ \ell\cdot\exp
\left(-\frac{1}{2}
\frac{ \frac{1}{2}mv^2 }{W} \right)
\ \ \ \ .
\end{eqnarray} 
The corresponding time is $T_{collision}=L_{collision}/v$. After 
that time the probability of being backscattered is of order 1.
In {\em one dimension} this backscattering will lead to (classical)
localization of the particle. The localization length is 
exponentially large for high energies. 

If $\tau_c$ is finite, rather than infinite, classical 
localization will manifest itself only if $T_{collision}\ll\tau_c$.
Within the time scale $\tau_c$, the particle spreads
over spatial range of the order $L_{collision}$, while 
its velocity is randomized. It follows that ${\cal K}(R|R_0)$
may be used as a stochastic kernel.  Hence the Markovian property 
is recovered over time scales that are much larger than $\tau_c$,
and a diffusive behavior follows with coefficient 
$L_{collision}^2/\tau_c$. This diffusion is non-dissipative 
``random walk'' like.


\section{Quantum Localization}
\setcounter{equation}{0}

As in the last subsection, we shall discuss the case of 
quenched disorder. However, here
the {\em quantal} analysis will be carried out.  
The variance of the disordered potential is $W^2$, with auto-correlation
length $\ell$. We approximate the Gaussian correlation (\ref{e2_6}) by 
a delta function. Defining $a=\sqrt{2\pi}\ell$, the path-integral 
expression for the propagator is (\ref{e6_7}) with (\ref{e6_10}) and
\begin{eqnarray}   \label{e11_1}
S_N[x',x''] = \frac{1}{2} aW^2 \int_0^t\int_0^t
(\delta(x_2''{-}x_1'')+\delta(x_2'{-}x_1')-2\delta(x_2''{-}x_1'))
d\tau_1 d\tau_2 
\end{eqnarray}
The essential feature of this functional, is its non-local
nature. Our first step will be to get some insight into $S_N$.

\begin{figure} 
\begin{center}
\leavevmode 
\epsfysize=6cm
\epsffile{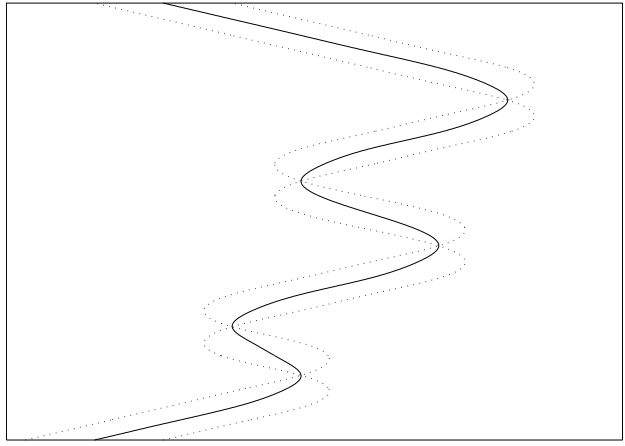}
\end{center}
\caption{\protect\footnotesize Illustration of an 
 optimal path $R$ (solid curve) for localization problem.
 The dotted lines are $x'(\tau)$ and $x''(\tau)$ that
 correspond to $r_R(\tau)$.    
 The horizontal axis represents spatial position, 
 while the vertical axis is for the time. }
\label{traj_loc}
\end{figure}

Let us consider segment $i$ and segment $j$ that belong both 
to the same path, either $x'$ or $x''$, within the same
spatial interval $\Delta x$. This includes the possibility $i=j$.
The contribution to $S_N$ is
\begin{eqnarray}
\Delta S_N = + \frac{1}{2} aW^2 \frac{\Delta x}{|v_i v_j|}
\nonumber
\end{eqnarray}
where $v$ is the velocity $\dot{x}$ within $\Delta x$. 
However, if $i$ belongs to $x'$, while $j$ belongs to $x''$,
or vise versa, then the contribution to $S_N$ is
\begin{eqnarray}
\Delta S_N = - aW^2 \frac{\Delta x}{|v_i v_j|}
\nonumber
\end{eqnarray}
One easily convince oneself that the contribution of
each spatial interval $\Delta x$ is {\em non-negative}. 
A zero value may be obtained if to each segment $i$, that 
belongs to $x'$, corresponds segment $j$ that belongs 
to $x''$, with $|v_j|=|v_i|$. For example, a zero value 
for $S_N$ may be obtained if $x''(\tau)$ is shifted in 
time with respect to $x'(\tau)$. 
Referring to the endpoints, one should identify $\tau=t$ 
with $\tau=0$. It follows that $S_N=0$ implies that the two paths
$x'$ and $x''$ are either {\em identical}, or satisfy  
the constraint $R(t)=R(0)$. If $R(t)\neq R(0)$, the value
$S_N=0$ may be obtained only if $x'$ and $x''$ are identical. 
More generally, if $|v_t|=|v_0|$, one may prove that the 
following inequality holds for any pair of smooth 
paths $x'$ and $x''$,  
\begin{eqnarray}    \label{e11_2}
e^{-\frac{1}{\hbar^2}S_N[R,r]} 
\ \ \le \ \ e^{-\frac{aW^2}{(\hbar v_0)^2}
\min(|R_t{-}R_0|,\frac{1}{2}(|r_t|{+}|r_0|))}
\ \ \ .
\end{eqnarray}
For particular $R(\tau)$ one may ask what is the $r(\tau)$ for
which $S_N[R,r]$ is minimal.  The trivial minimum, which is 
also the absolute minimum, is $r(\tau)=0$, for which
$S_N[R,r]=0$. However, any small perturbation on $r(\tau)$
will make $S_N[R,r]$ much larger.  Therefore, we are tempted 
to assume that there may be some other, more stable (local) 
minimum $r_R(\tau)$. A non-trivial local minimum $r_R(\tau)$ does 
not exist for any $R(\tau)$. However, one can prove that there is a 
large family of $R$-s for which such minimum exists,  
by actually constructing them. This is done by following the 
considerations that were presented at the beginning of this 
paragraph, An example for such construction is presented 
in Fig.\ref{traj_loc}.
The situation here should be contrasted with 
that encountered in Sec.VIII. In subsection A (there)
we could have defined {\em one} optimal path $R_o(\tau)$. 
Here, there is a whole family of ``optimal paths'', as in 
the case of subsection B (there). However, in the present
case these paths are ``interference paths'' rather than 
``classical paths''. The following observations concerning 
the relevant optimal paths are important: (a) They consist of 
many straight segments, and have ZigZag character; (b)  The final 
velocity $\dot{R}(t)$ is favored to be equal in absolute value to 
the initial velocity; (c) Turning points impose significant 
restrictions; (d) The non trivial minimum $r_R(\tau)$ is  
isolated; (e)  The non-trivial minimum is relatively stable.     
The last point is the most difficult to observe. First it should 
be noted that (e) must be true a-priori. Else, if the trivial minimum 
dominates the path-integral expression, then the result would be 
that the particle has roughly the same probability to go 
from any initial conditions to any final conditions, irrespective 
of proximity considerations or even energy conservation. 
Still, a reasonable argument is required why the non trivial 
minimum is relatively stable.  For this consider a straight 
segment $i$ for which $\dot{R}(\tau)=v_i$ and $r_R(\tau)=r_i$.  
One observes that if $r_R$ is perturbed by a fluctuation of time period 
$|r_i|/|v|$, or by some higher harmony, then the contribution to 
$S_N[R,r]$ is negligible.  This is to be contrasted with the case 
of the trivial minimum $r(\tau)=0$, where any fluctuation has
high cost.

We turn now to the formal extension of the procedure 
that has been presented in the previous section. 
We expand $S_N[R,r]$ around the non-trivial minimum:
\begin{eqnarray}   \label{e11_3}
S_N[R,r] \approx S_N[R,r_R] +
\int_0^{\infty}\int_0^{\infty}d\tau d\tau' \ 
\phi_R(\tau{-}\tau') \cdot (r(\tau){-}r_R(\tau)) 
\cdot (r(\tau'){-}r_R(\tau')) \ + \ ... 
\end{eqnarray}
Here we are not able to write an explicit expression for 
the highly complicated kernel $\phi_R(\tau{-}\tau')$. However,
we proceed, and write down the result of the Gaussian 
integration:
\begin{eqnarray}   \label{e11_4}
{\cal K}[R] \ = \ \ const \cdot
\cos\left(\int_0^{\infty}d\tau \ \Delta_R(\tau)r_R(\tau)\right)
\ \mbox{e}^{-S_N[R,r_R]} \ \ \cdot \nonumber \\  
\cdot \ \ \exp\left(-\frac{1}{2}\int_0^t\int_0^t
d\tau d\tau' \ \Phi_R(\tau-\tau') 
\ \Delta_R(\tau) \ \Delta_R(\tau')\right)
\end{eqnarray}
Now, we should perform the ${\cal D}R$ integration.  This 
integration will be dominated by the family of ``optimal
paths''. Note that the $cosine$ term in (\ref{e11_4}) equals unity
since for $r_R$ the trajectories $x'$ and $x''$ are in 
a sense ``shifted'' one with respect to the other, hence 
$S_{eff}=0$. Within the family of optimal paths, not 
all have the same contribution. One should expand around 
those that have the largest contribution. For these paths
$S_N[R,r]=\frac{aW^2}{v_0^2}|R-R_0|$. Here we consider 
endpoints conditions $(R_0,mv_0)$ at $t=0$ and 
$(R,\pm mv_0)$ at time $t$. The time $t$ is assumed to be 
sufficiently large to guaranty steady state distribution.
It follows that   
\begin{eqnarray}   \label{e11_5}
{\cal K}(R,\pm mv_0|R_0,mv_0) \ \sim \ \ 
\exp\left( -2 \frac{|R-R_0|}{\xi(v_0)} \right)
\end{eqnarray}
where
\begin{eqnarray}   \label{e11_6}
\xi(v_0)=2\cdot \frac{(\hbar v_0)^2}{aW^2} \ \ \ .
\end{eqnarray}
For convenience $\hbar$ has been restored in the
latter formula.  
In the application of the inequality (\ref{e11_2}) 
a sub-family of ``optimal paths'' has been ignored, for
which $|r_R(endpoints)|<|R{-}R_0|$. It is justified
provided this sub-family constitutes a zero fraction 
of the whole family.  For this we should assume 
sufficiently long time ($t$), for which a steady 
state distribution is attained.  

Both the {\em validity} and the {\em applicability} of 
the inequality (\ref{e11_2}) demonstrates the 
vulnerability of our localization argument. It is important to 
consider circumstances in which either of these conditions 
is not satisfied. The inequality (\ref{e11_2}) will not be
{\em valid} if the noise is not infinitely correlated in 
time, i.e. the disordered potential is not static when 
viewed on large time scales.  The inequality (\ref{e11_2})
will not be {\em applicable} if additional white noise 
is added to the Hamiltonian. In the latter case, 
large $r_R(\tau)$ will be suppressed by the corresponding 
additional term in the noise functional. Consequently, 
the sub-family of paths for which $|r_R(endpoints)|<|R{-}R_0|$
will not constitute zero fraction of the whole family.

Expression (\ref{e11_6}) for the localization length 
agrees with the well know result for 1D localization
of ``free'' particle (Thouless Ref.\cite{thouless}). Note that Thouless uses
some scaled units, resulting in the expression 
$\xi=8mE/\hbar^2$. We shall show now that the above expression 
is somewhat more general, and applies also to cases
where the dispersion law is different from $v=p/m$. 
For this one should replace in (\ref{e6_9}) the kinetic
term  $\frac{1}{2}m\dot{x}^2$ by some general function $T(\dot{x})$, 
resulting in
\begin{eqnarray}    \label{e11_7}
S_{free}\ = \ \int_0^t d\tau (T(\dot{x}''){-}T(\dot{x}')) \ 
\approx \ \int_0^t d\tau \ p(\dot{R}) \ \dot{r} \ + ...
\end{eqnarray}
where $p(v)=T'(v)$ is the dispersion law. The subsequent
formalism is easily generalized. The expression (\ref{e11_6})
for $\xi$ is unaffected. 
Actually, the general expression for $\xi$ could have been 
guessed. We use naively the Born approximation, calculate the 
mean free path $\ell$, and rely on the fact that $\xi$ is twice 
the mean free path (Thouless Ref.\cite{thouless}). By the golden rule, 
the probability of being backscattered is
\begin{eqnarray}  \label{e11_8}
Prob=2\pi\overline{| \langle -p|{\cal U}|p \rangle |^2}\cdot
\frac{L}{2\pi}\frac{dp}{dE}
\end{eqnarray}   
where $L$ is the length of the available space, and $E$ is
the kinetic energy. The matrix element is
\begin{eqnarray}   \label{e11_9}
\overline{| \langle -p|{\cal U}|p \rangle |^2}=\frac{1}{L^2}\cdot 
\overline{|\mbox{FT}[{\cal U}]|^2} = \frac{1}{L} aW^2
\end{eqnarray}
where in the last equality FT[] denotes Fourier Transform, 
and ${\cal U}$ is assumed to be
uncorrelated in space (``white spatial noise''). The result 
(\ref{e11_6}) is easily recovered. One should use $\xi=2\ell$,
and $v_0=dE/dp$, while $\ell$ is evaluated via the relation 
$Prob=v_0/\ell$. 
In order to further demonstrate the generality of (\ref{e11_6}),
let us consider Anderson tight binding model. The spacing
will be denoted by $a$. The Hamiltonian is \linebreak
\mbox{${\cal H}=\sum_{n} (|n\rangle V_n\langle n| 
+ T(|n\rangle\langle n{-}1|+|n\rangle\langle n{+}1|))$}
where $V_n$ is uniformly distributed in $[-W_0,W_0]$. The 
transition amplitude is $T$. The Kinetic energy in the 
center of the band is $E=2Tap$, where $p$ is the momentum.
Hence $v_0=2Ta$ there. The dispersion of the on-site energies
is $W=\frac{1}{\sqrt{12}}2W_0$. The ``area'' under the triangular
autocorrelation function is $aW^2$. Formula (\ref{e11_6}) suggests 
that $\xi=24a(T/W_0)^2$. This results agrees with that of
Ref.\cite{kramer} (equation (66) there), including the prefactor.


\section{Summary and Conclusions}
\setcounter{equation}{0}

A unified treatment of Diffusion Localization and Dissipation (DLD) 
has been presented in this work. All these phenomena may 
be derived from the general path integral expression (\ref{e6_7}),
\begin{eqnarray}  
{\cal K}(R,P|R_0,P_0)=
\int_{R_0,P_0}^{R,P} \!\!\! {\cal D}R \int {\cal D}r 
\ \ \mbox{e}^{i\frac{1}{\hbar}S_{eff}[R,r]} 
\ \ \mbox{e}^{-\frac{1}{\hbar^2}S_N[R,r]}
\ \ \ \ ,
\end{eqnarray}
upon inclusion of the appropriate functionals $S_{eff}$ and $S_N$. 
General expressions for these functionals are available and various 
limits may be considered: (a) Quantal versus classical expressions; 
(b) Disordered versus non-disordered environment; (c) Dissipative 
versus non-dissipative environment;  (d) Quenched versus noisy 
environment.  In the classical limit the DLD model constitutes 
a formal solution of Langevin equation.  The classical limit may 
be obtained by linearization of the quantal $S_{eff}[R,r]$ with 
respect to $r$, while expanding $S_N[R,r]$ to be quadratic in 
this path-variable. The disorder or its absence depends on the choice 
of the spatial auto-correlation function $w(r)$.  The dissipation 
is turned on if the friction functional $S_F$ is included in  
$S_{eff}$. The nature of the noise, whether it is ``quenched'', ``colored''
or ``white'', is determined by the noise kernel $\phi(\tau)$. In 
the latter case any combination may be considered as well (see further 
discussion at the end of this section).

The classical BM model is well defined in terms of an 
appropriate Langevin equation only in the case of an ohmic bath.
This is not the case with the classical DLD model.
The classical dynamics in the latter case is well defined 
in terms of appropriate Langevin equation also for 
non-ohmic bath. Explicit expressions for 
the friction force, and for the effective mass have 
been derived. Another nice feature of the DLD model is
the absence of ``switching impulse''.

Once the noise auto-correlation function $\phi(\tau)$ is specified, 
the BM model is in-distinguishable from its classical limit. 
As long as the external potential $V(x)$ is quadratic (at most), 
the quantal propagator is identical with the classical one, and 
Langevin equation can be used in order to describe correctly 
the time evolution of Wigner function. All the quantal effects that 
are associated with the standard Zwanzig-Caldeira-Leggett BM-model 
are (formally) reproduced by solving the classical Langevin equation 
with an appropriate noise term. {\em The DLD model is different}. 
The non-stochastic, genuine quantal features of the DLD model have 
been discussed. These features constitute a manifestation of either
the disorder or the chaotic nature of the bath.

For either noisy or ohmic environment both the BM model and the 
DLD model leads to either spreading or diffusion.  For the BM model, 
the spreading and the diffusion profiles are described by 
a Gaussian distribution. For the DLD model one should include 
a quantal correction. The disorder freezes a piece of the wavepacket, 
letting it to propagate as if it were a free particle.  
Both, the ``quantal correction'' to the propagator, as well as 
any other interference phenomena, die out exponentially in time.  
This exponential decay, due to dephasing, 
is independent of geometry. {\em It should be contrasted} with the 
results for loss of interference in the presence of BM-like environment
\cite{AAK,stern}. Another important observation is 
that for {\em disordered} environment, quantal interference 
is unaffected by friction. 

On the classical level it is fascinating to analyze the 
diffusion profile in the presence of disorder.  For the 
low temperature ohmic BM-model, it is found that diffusion is
suppressed, though its Gaussian profile is maintained. 
The DLD model, in the same circumstances, leads to an 
exponential profile that does not change with time. 
This new effect is due to the interplay of the 
temporal (negative) autocorrelations of the noise with the 
spatial disorder. Even more fascinating ``diffusion'' profiles 
are found for other types of noise autocorrelations. 

Quenched disorder in {\em one dimension} leads to classical 
localization, as well as to quantal localization. The former 
is characterized by a localization length which is exponentially
large at high energies. Quantal localization on the other hand 
is dominated by interference phenomenon. We have identified 
the interference paths within the framework of FV formalism, 
and demonstrated how the well known exponential profile emerges. 
The localization length in the quantal case is proportional to 
the square of the velocity. The applicability of this result to 
other dispersion relations has been pointed out.   
 
It is obvious from the derivation, that localization 
cannot be argued if the noise is not strictly static
(e.g. slowly  modulated). Diffusive behavior is recovered 
if the noise possess long but {\em finite} auto-correlation time.    
Also the case of white noise ``on top'' of the static 
disorder will evidently lead to diffusion \cite{ott}\cite{noise}.  
In this latter case, which has not been considered in this paper,  
non perturbative effects may manifest themselves \cite{shep} as 
in the case of the ``Quantum Kicked Particle'' \cite{noise}. 
The DLD model should also account for diffusion in 
the presence of both quenched disorder, noisy potential 
and friction (all together).  There should 
be a way to derive systematically well known 
heuristic results that corresponds to hopping, or 
variable range hopping \cite{mott}.     
   
Classical non-dissipative ``random walk'' diffusion has been 
discussed in the restricted case of long-time correlated noise.   
It is essential to generalize the DLD model to 
{\em more than one dimension} in order to account for this 
phenomenon in the case of strictly static disorder. 
Auto-correlations of the disordered potential, both in time 
and in space, should be considered, in order to generate the 
dynamics which is described by Boltzmann transport equation. 
In the limit of quenched disorder, localization effect should 
be encountered.

\ \\
\noindent ACKNOWLEDGMENT

\noindent I thank Shmuel Fishman from the Technion for supporting 
and encouraging the present study in its initial phases.  I also thank 
Harel Primack and Uzy Smilansky for interesting discussions and suggestions, 
and Yuval Gefen for his comments. 
The research reported here was supported in part by  the Minerva Center 
for Nonlinear Physics of Complex systems.
 

\setcounter{section}{0}
\def\thesection{\Alph{section}}
\def\theequation{\Alph{section}.\arabic{equation}}

\section{Interaction with External Bath that
           Consists of Extended Field Modes}  
\setcounter{equation}{0}

This appendix illustrates how the unified formalism that has been 
presented in this paper should be modified in order to deal with
external bath that consists of extended field modes. Unlike the 
case of the BM-model, this modification is not as immediate as 
the mere substitution of appropriate $w(r)$.
In the present derivation the bath Hamiltonian is not specified, 
and also the assumption $u_{\alpha}=u(x{-}x_{\alpha})$ is altered.

For the standard derivation of the DLD model it has been assumed that 
the interaction Hamiltonian is (\ref{e3_2}) with 
$u_{\alpha}=u(x{-}x_{\alpha})$, leading to the factorized noise 
auto-correlation function (\ref{e2_4}). In the more general case 
the classical derivation as well as the quantal derivations 
lead to a stochastic force that satisfies (\ref{e2_3}) or to 
the noise functional (\ref{e6_11}) respectively, with   
noise auto-correlation function which is 
\begin{eqnarray} \label{eA_phi}  
\phi(x{-}x',t{-}t')=\sum_{\alpha}\phi_{\alpha}(t{-}t') 
\ u_{\alpha}(x)u_{\alpha}(x') \ \ \ .
\end{eqnarray}
Here 
$\phi_{\alpha}(t{-}t'){=}\langle Q_{\alpha}(t)Q_{\alpha}(t')\rangle_{eq}$ \ ,  
and the interaction Hamiltonian (\ref{e3_2}) is used
with $c_{\alpha}{=}1$ without loss of generality.
The fields mode are still assumed to be decoupled, 
but the bath Hamiltonian is not specified. Instead, 
one relys on Fluctuation-Dissipation theorem as in 
Ref.\cite{stern}, writing the general expression: 
\begin{eqnarray}  
\phi_{\alpha}(t{-}t') \ = \ \int_0^{\infty}\frac{d\omega}{\pi}
J_{\alpha}(\omega) \ \hbar\coth(\frac{1}{2}\hbar\beta\omega)
\ \cos[\omega(t{-}t')] \ \ \ \ .
\end{eqnarray}
Note that if the field modes are simple harmonic oscillators,
then one should substitute 
$J_{\alpha}(\omega){=}\frac{\pi}{2m_{\alpha}\omega_{\alpha}}
\delta(\omega{-}\omega_{\alpha})$.  Alternatively, one may 
speculate $J_{\alpha}(\omega)$ using known response characteristics
of the bath.

To make further progress the interaction Hamiltonian 
(\ref{e3_2}) should be further specified. Using ``standing waves''
decomposition it is assumed to be 
\begin{eqnarray} 
{\cal H}_{int}\ = \ \sum_{\alpha} \left(
Q_{1\alpha}\cos(\vec{k}_{\alpha}\cdot x) +
Q_{2\alpha}\sin(\vec{k}_{\alpha}\cdot x) \right)
\ \ \ \ .
\end{eqnarray}
Substitution of the appropriate $u_{\alpha}$ into (\ref{eA_phi}),
and converting the summation to an integral over directions and 
over $k$, the result can be cast into the following form
\begin{eqnarray}
\phi(x{-}x',t{-}t') \ = \ \int_0^{\infty}\frac{k^{d{-}1}dk}{\pi}
\phi(k,t{-}t') \ \langle\cos(k\hat{\Omega}\cdot(x{-}x'))
\rangle_{\Omega}  \ \ \ \ . 
\end{eqnarray}
The average over directions can be performed leading
to $\cos(k|x{-}x'|)$ in 1-D, Bessel function ${\cal J}_0(k|x{-}x'|)$ 
in 2-D, and $\mbox{sinc}(k|x{-}x'|)$ in 3-D. 

In Ref.\cite{stern} the interaction with electromagnetic fluctuations 
in metal has been considered. It has been assumed that each 
mode is characterized by an ohmic response. In our notations 
it corresponds to $J(k,\omega)=\eta(k){\cdot}\omega$. Due to this 
assumption, the noise auto-correlation function becomes factorized
at high temperatures, as in the DLD model. In particular, for 
the noise functional (\ref{e6_11}) one obtains
\begin{eqnarray}
S_N[r] \ = \ \int_0^t d\tau \ \frac{1}{\beta}
\int_0^{\infty}\frac{k^{d{-}1}dk}{\pi}
\ \eta(k) \ \langle\sin^2(k\hat{\Omega}\cdot r(\tau))
\rangle_{\Omega}  \ \ \ \ . 
\end{eqnarray}    
For the discussion of dephasing, we write again this result
using the notations of Ref.\cite{stern}. It is  assumed that 
$\eta(k)=e^2/(\sigma k^2)$, where $e$ is the charge of the 
electron and $\sigma$ is the conductivity.  Integration over 
$k$ leads to the result
\begin{eqnarray} \label{eA_supp1}   
\langle \mbox{e}^{i\varphi} \rangle \  = \ \
\exp\left[-\frac{e^2k_B T}{2\hbar^2\sigma}\int_0^t 
\ |x_a(\tau)-x_b(\tau)|^{2-d} \ d\tau \right]  \ \ \ \ 
\mbox{for $d \le 2$} \ . 
\end{eqnarray}
This result, that has been obtained in Ref.\cite{stern} 
(equation (7.8) there), is very similar to the corresponding 
result (\ref{e_supp}) for the BM-model, the difference being 
the power $(2{-}d)$ which is not universal.   

An apparently simpler example for an interaction with external 
bath that consists of extended field modes, is the 
electron-phonon interaction. Here we want to question 
the applicability of the BM model as an approximated description. 
The electron is assumed to be confined to a one dimensional 
``quantum wire'', while the phonons dwell in the 3-D bulk. 
Considering longitudinal modes, the coupling of the 
$\alpha \rightarrow (k_1,k_2,k_3)$ ``oscillator'' with the 
electron is $k_1 \cdot x$, with a coupling constant 
$C_{\alpha}\propto\sqrt{k_1^2+k_2^2+k_3^2}$. 
Summing over $\alpha$, as defined in (\ref{e3_5}), one obtains
effectively $J(\omega)\propto \omega^{d+2}$ where $d{=}3$ is the 
dimensionality of the space where the phonons dwell.
Thus a phonons-bath is similar to a superohmic BM-model. 
However, this derivation is somewhat miss-leading, since 
a cutoff $\omega_c$ should be introduced, while for the 
original electron-phonon interaction a natural 
cutoff $|x{-}x'|/c$ exists.


\section{An Interference Gedanken Experiment}
\setcounter{equation}{0}

In this appendix we consider the interference phenomenon from two different
points of view, and demonstrate their consistency. First we consider the free
propagation of two-wavepackets superposition. The decay of the interference 
pattern will be dictated by the propagator (\ref{e9_2}). Then we consider the 
scattering of a simple wavepacket from a double-barrier. The suppression 
of interference paths will be dictated by the noise functional (Sec.VI(C)).
Finally, we argue that both points of view are physically equivalent, and 
must lead to the {\em same} result, which is indeed the case.

Consider a superposition of two Gaussian wavepackets which have the same
momentum $P_0$, the same initial spatial spreading $\sigma_0$, while 
their initial locations satisfies $|R_{02}{-}R_{01}|{=}d$. The Wigner 
function for this preparation is easily computed, and is of the form
\begin{eqnarray}  \nonumber  
\rho_{t{=}0}(R,P)\ = \ \frac{1}{2}G(R{-}R_{01},P{-}P_0)
\ + \ \frac{1}{2}G(R{-}R_{01},P{-}P_0) \ + \ 
\cos\left(\frac{P-P_0}{\delta P_c}\right)\cdot
G(R{-}\frac{1}{2}(R_{01}{+}R_{02}),P{-}P_0) \ \ \ \ , 
\end{eqnarray}
where $G(x,p)$ denotes ``minimum-uncertainty'' Gaussian distribution,
and $\delta P_c{=}\hbar/d$. For free propagation $\rho_t(R,P)$ will
develop fringes on the spatial scale 
$\Delta x{=}\frac{\delta P_c}{m}{\cdot}t \ $.
Note that this gedanken experiment is formally equivalent to the usual
``two slit'' diffraction experiment upon the definition 
$\Delta\theta=\frac{\Delta x}{v{\cdot}t}=\frac{\hbar}{P_0}/d$.
For propagation in noisy non-disordered environment the interference 
pattern is smeared on scale $\delta P\sim\nu{\cdot}t$, due to the diffusive 
momentum spreading. The smearing factor is 
$\exp[-\frac{1}{2}(\delta P(t)/\delta P_c)^2]$ 
leading to an exponential decay $\exp[-\frac{\nu d^2}{\hbar^2}t]$, 
that depends on the separation $d$. 
On the other hand, for propagation in noisy 
disordered environment, using (\ref{e9_2}), the 
exponential decay is $\exp[-\frac{\nu\ell^2}{\hbar^2}t]$, 
independent of geometry.

Consider now scattering problem in one dimension. The potential 
is assumed to be $V(x){=}\infty$ for $x<0$, and 
$V(x)=\delta(x{-}\frac{d}{2})$ for $0<x$. A simple Gaussian 
wavepacket with momentum $-P_0$ and spatial spreading $\sigma_0\ll d$
is launched from $R_0$.  The scattered particle is detected 
at the range $R$ which is assumed to be much larger than $R_0$.
In the absence of noise and dissipation, it is not difficult to work 
out the explicit solution of this scattering problem. A ``train''
of Gaussian-like wavepackets will emerge from the scattering region.
The spatial separation between each two wavepackets is $d$, and the 
probability density function will contain interference pattern 
in between. From the FV path-integral point of view (Sec.VI(C)), 
the interference pattern is due to the existence of 
``interference paths''. In the presence of noisy disordered environment 
the contribution of these interference paths to the propagator
is suppressed exponentially in time. 

Assuming that the dephasing during the time interval 
$t_{scattering}{=}R_0/v$ is negligible, one may use the results 
of the first gedanken experiment, leading to the same 
exponential decay. Thus, the result (\ref{e9_2}) for the 
propagator is consistent with the analysis of interference 
in Sec.VI(C). Our gedanken experiment can be used also in order 
to illustrate and clarify the observation that friction does 
not affect the interference phenomenon. This observation holds for 
disordered dynamical environment.


\end{document}